# Stripes and the  (Cu)$_{13}$-BEC model

## A. Rosencwaig

*Department of Physics, University of Toronto, Toronto, Ontario M5S 1A7, Canada*

## ABSTRACT

The (Cu)$_{13}$-BEC model of high-temperature superconductivity was previously shown  to account for many of the principal thermodynamic and electronic properties of the superconducting cuprates. Here I show that this model is also able to account for many of the key characteristics of the coupled CDW and SDW orders in these compounds. These include the general coexistence of metallic parallel stripes with superconductivity, the well-known linear relationship between the incommensurability of the SDW-induced IC magnetic peaks and the dopant concentration, as well as the saturation of this incommensurabilty at $\approx 1/8$ for doping $\geq 1/8$. The model also provides a natural explanation for the celebrated 1/8-anomaly in LSCO and LBCO. It is also able to account for the severe suppression of the superconductivity in LNSCO at all doping levels and for the variations in the properties of LBSCO, at a fixed hole doping of 1/8, as its crystal structure is changed. Furthermore, the (Cu)$_{13}$-BEC model is also consistent with many of the characteristics of the SDW orders in Y123. Most importantly, scanning tunneling microscopy results on Bi2212 appear to provide a direct validation of the CDW order predicted by the model.

Correspondence should be addressed to the author at e-mail:
allanrosencwaig@hotmail.com

## 1. INTRODUCTION

High-temperature cuprate superconductors exhibit many interesting thermodynamic, electronic and magnetic properties. In the last few years considerable interest has focused on the unusual magnetic properties wherein static and dynamic charge-density-wave, CDW, and spin-density-wave, SDW, orders give rise to static and fluctuating charge and spin stripe patterns that at times appear to coexist and at other times appear to compete with superconductivity.[1-4] A fundamental question, therefore, concerns the relationship between these spin-charge orders and high-temperature superconductivity.

Charge stripes result from the competition between the kinetic energy of mobile dopant charges and the superexchange interaction between neighboring $Cu^{+2}$ spins.[5-13] For hole-doped cuprates this competition appears to result, at low temperatures, in a spatial segregation of holes into hole-rich one-dimensional regions which form antiphase domain walls between hole-poor regions of antiferromagnetically correlated $Cu^{+2}$ spins. The localization of the holes in the domain walls increases the kinetic energy of the holes, while decreasing the potential energy of the $Cu^{+2}$ spins, resulting in a net decrease of total energy. Stripe theories predict that charge stripes can be oriented at 45° to the Cu-O-Cu bonds, now referred to as "diagonal" stripes, or parallel to the Cu-O-Cu bonds, now referred to as "parallel" stripes, each with a charge density along the stripe of $\approx 1$ charge /Cu, and further that the stripes are insulating. Indeed, the nickelates, $La_2NiO_{4.125}$ and $La_{1.8}Sr_{0.2}NiO_4$, do exhibit insulating diagonal charge stripes with a charge density of $\approx 1$ hole/Ni.[14, 15]

The cuprates exhibit a much more complex behavior. At very low doping, insulating diagonal stripes with a charge density of $\approx 1$ hole/Cu are observed, similar to the case in the nickelates.[16-19] However, with the onset of superconductivity, parallel stripes begin to appear and to coexist with the superconductivity order.[16, 19] Moreover, these stripes have a charge density of $\approx 0.5$ holes/Cu and are metallic. In some cuprates, notably $La_{2-x}Sr_xCuO_4$, LSCO, and its related compounds such as, $La_{1.875}Ba_{0.125}CuO_4$, LBCO, and $La_{1.6-x}Nd_{0.4}Sr_xCuO_4$, LNSCO, the presence of stripes appears at times to compete with superconductivity. This is especially true in LNSCO where superconductivity is severely suppressed throughout the superconducting doping range.[20, 21] In LSCO and LBCO, superconductivity is also suppressed, mildly in LSCO and severely in LBCO, near the "magic" dopant concentration of 1/8, the celebrated "1/8- anomaly".[22, 23]

I have recently proposed a Bose-Einstein condensation model, the $(Cu)_{13}$-BEC model, for high-temperature superconductivity that appears to account for many of the principal thermodynamic and electronic properties of the cuprates.[24] The model accounts for the basic bell-shaped $T_c$ vs doping curve, the $T_c$ dependence on the number of $CuO_2$ layers, and for many of the key experimental angle-resolved photoemission spectroscopy, ARPES, muon spin resonance, μSR, and microwave results on the temperature and doping dependencies of both the superfluid density and the pairing strengths (superconducting gap, leading-edge-midpoint and psuedogap) in these high-temperature superconductors. Here I show that this same $(Cu)_{13}$-BEC model is consistent with many of the key magnetic properties, that is, with the key charge-spin stripe properties of the cuprates as well.

## 2. THE $(Cu)_{13}$ CLUSTERS

In the $(Cu)_{13}$-BEC model, direct Cu-Cu charge hopping up to third nearest neighbor Cu results in a dopant singlet bonding state arising from the hybridization of dopant charge anti-bonding molecular orbitals in a cluster containing 13 Cu and 26 O atoms in the $CuO_2$ plane of the superconducting cuprates. This singlet $(\psi_o)^2$ state forms a pre-formed charge pair with relatively low binding energy ( typically < 5 meV), that is spatially bounded, with a wavelength or interaction length $\lambda \approx 5.5a_t$, where $a_t$ is the tetragonal lattice constant. This singlet cluster state is also weakly interacting because of screening from the potential of the underlying dopant ions. In the model we assume that the $(\psi_o)^2$ singlet state retains enough phase coherence during transport through the lattice that it acts as a bosonic quasiparticle, and thus is capable of undergoing Bose-Einstein condensation. The bosonic singlet state is formed from a linear combination of 13 anti-bonding states of the form $\sigma^*_{x^2-y^2} = d_{x^2-y^2} -_+ p\sigma_x \pm p\sigma_y$. The hybridized $(\psi_o)^2$ state has a predominantly $d_{x^2-y^2}$ or $d_{-(x^2-y^2)}$ symmetry with the $p\sigma_x$ and $p\sigma_y$ orbitals averaging out. As depicted in Fig. 1, the $(Cu)_{13}$ cluster has an internal charge distribution that is evenly distributed along the x and y axes, that is, parallel to the Cu-O-Cu bonds and thus parallel to the tetragonal $a_t$ and $b_t$ axes. The cluster has $d_{x^2-y^2}$ or $d_{-(x^2-y^2)}$ symmetry for the $(\psi_o)^2$ state when the anti-bonding state at the central Cu site is $\sigma^*_{x^2-y^2}$ or $\sigma^*_{-(x^2-y^2)}$ respectively. Since the phase of the anti-bonding state alternates between neighboring Cu sites, the cluster phase will alternate between $d_{x^2-y^2}$ and $d_{-(x^2-y^2)}$ symmetry in the same manner.

The basic Bose-Einstein condensation condition is given by,

$$n_b\lambda^3 = 2.612 \qquad\qquad (1)$$

where $n_b$ is the boson density and $\lambda$ is the wavelength or interaction distance (as a diameter) of the boson, and Eqn (1) simply states that Bose-Einstein condensation, BEC, occurs when the average distance between bosons becomes somewhat less than the wavelength or interaction distance. In conventional BEC,[25] one assumes that the boson density, $n_b$, is independent of temperature, T, while $\lambda$ is the thermal wavelength and is thus dependent on T. In the $(Cu)_{13}$-BEC model we assume the opposite. The interaction distance, $\lambda$, is set primarily by the cluster geometry. As seen in Fig. 1, the $(Cu)_{13}$ clusters have a minimum diameter of $4a_t$. We assume that charge hopping between clusters can occur whenever an outer Cu of one cluster is at a distance of one $a_t$ (nearest-neighbor), $\sqrt{2}a_t$ (next-nearest neighbor) or $2a_t$ (third-nearest neighbor) from an outer Cu atom of the other cluster. In the model we thus assume an average $\lambda \approx 5.5a_t$, essentially independent of T. However, the density of the pre-formed pairs, or bosonic quasiparticles, is highly dependent on T since it depends on the Botzmann population density of the singlet $(\psi_o)^2$ state.

We then find that the basic BEC condition for the superconducting cuprates is,[24]

$$\frac{1}{2} nqP(\delta,T) = 2.612 \, a_t^2 c / \lambda^3 \qquad (2)$$

where n is the number of $CuO_2$ layers in the primitive tetragonal unit cell of dimension $(a_t \times a_t \times c)$ $A^3$, q is the dopant charge per Cu atom in the $CuO_2$ plane, and $P(\delta,T)$ is the probability that a charge is in the $\psi_0$ ground state at temperature T, with an energy gap $\delta$ separating $\psi_0$ from the next state. For a simple cluster energy diagram of 13 non-degenerate states all separated from each other by the same energy gap $\delta$, we have,

$$P(\delta,T) = 1/[1 + e^{-\delta/kT} + e^{-2\delta/kT} + \cdots\cdots + e^{-12\delta/kT}] \qquad (3)$$

Since $13qP(\delta,T)$ represents the number of charges in the $\psi_0$ state per $CuO_2$ layer in a cluster, and since there is only one $\psi_0$ state per layer in each cluster, a second BEC condition is,

$$13qP(\delta,T) \leq 2 \qquad (4)$$

The superconducting transition temperature, $T_c$, is obtained by solving Eqn. (2) for $T = T_c$ at a given q and $\delta$. The energy gap between the bosonic ground state and the next higher-energy state is is given by $\delta$. Thus $\delta$ is the binding energy of the pre-formed pair, while the superconducting pairing strength or superconducting gap is given by $12\delta$. In the $(Cu)_{13}$-BEC model $\delta$ is found to decrease linearly with q throughout the superconductivity doping range (typically $0.05 < q < 0.27$).[24] Eqn. (2) then results in the typical bell-shaped $T_c$ vs q curve, where $T_c$ first increases with q since the density of the bosonic quasiparticles, or superfluid density, initially increases with q. At $q = 2/13$, the cluster has, at $T = 0$, two charges in the $(\psi_o)^2$ state, or one bosonic quasiparticle, the maximum possible per $CuO_2$ layer in a cluster, as given by Eqn. (4). Thus the $T = 0$ superfluid density reaches a maximum and saturates at $q = 2/13$. However, the pairing strength $\delta$ continues to decrease with q and this causes $T_c$ to peak in the vicinity of $q \approx 2/13$ and then decrease for $q > 2/13$. Finally $T_c \to 0$ and superconductivity ceases when $\delta \to 0$ at $q \approx 0.26$ - $0.27$. Note that the calculated $T_c$ depends on the magnitude of the bosonic quasiparticle or superfluid density. We shall see later that it is possible for charge pinning to perturb the $(\psi_o)^2$ state and to reduce this density, thereby reducing $T_c$.

### 3. CLUSTER ORDERING

Stripe models assume that charge is ordered within one-dimensional domain walls, although some models also provide for a two-dimensional checkerboard pattern as well.[12] Here we will assume that, once the $(Cu)_{13}$ clusters begin to form, the physics that gives rise to charge ordering can also produce cluster ordering. This will occur for the following reason. According to the $(Cu)_{13}$-BEC model, clusters are weakly interacting because of the screening potential from the underlying dopant ions. However, this is true only if the clusters are randomly distributed and if they have a minimum separation of at least one $a_t$. The ordering of clusters into stripes results in a non-random distribution of clusters and, as we shall see later, in direct contact of clusters along the stripe direction. Under these conditions Coulomb repulsion effects are important and, as in the case of charge stripes, these effects can lead to an increased kinetic energy of the two charges in the cluster. However, this increase in hole kinetic energy is compensated for by an even greater decrease in the spin potential energy of the antiferromagnetically correlated regions between the cluster stripes. Thus cluster stripes, like charge stripes, can result in hole-rich antiphase domain walls separating hole-poor antiferromagnetically correlated regions. The conclusion is that $(Cu)_{13}$ clusters can also order into stripe patterns due to the competition between charge kinetic energy and spin potential energy.

It should be noted that, in the $(Cu)_{13}$-BEC model, ordering of the $(Cu)_{13}$ clusters does not, by itself, affect the superconductivity as long as the $(\psi_o)^2$ state remains essentially unperturbed and the superfluid density and pairing strength remain unchanged. Thus, according to the model, cluster ordering and superconductivity can coexist, although as we will see later, they can, at times, affect each other. Applying the model and the concept of cluster ordering, let us now see how we would expect stripes to evolve in LSCO as a function of hole doping.

## 4. LSCO: DIAGONAL STRIPES

First, let us discuss briefly how CDW and SDW orders are detected in the cuprates. Stripe models of coupled CDW and SDW orders predict that there will be CDW-induced superlattice diffraction peaks split by $(\pm 2\varepsilon, 0, 0)$ about some fundamental reciprocal lattice point (h, k, l) for CDW stripes aligned along k. A similar set of peaks will be split by $(0, \pm 2\varepsilon, 0)$ for CDW stripes along h. The incommensurability of the CDW-induced peaks is $2\varepsilon$, and the splitting between the peaks is $4\varepsilon$. Usually $\varepsilon$ is given in terms of reciprocal tetragonal lattice units, $a_t^{-1}$. Since the charge stripes form domain walls that separate antiphase antiferromagnetic domains, the modulation period of the magnetic superlattice structure is twice that of the charge stripes. Thus the SDW-induced superlattice diffraction peaks will be split by $(\pm\varepsilon, 0, 0)$, or $(0, \pm\varepsilon, 0)$, about the appropriate fundamental reciprocal magnetic lattice point. The incommensurability of the SDW-induced superlattice diffraction peaks is thus $\varepsilon$, and the splitting between peaks is $2\varepsilon$.

Although SDW-induced superlattice diffraction peaks can be readily detected by neutron scattering, CDW-induced superlattice diffraction peaks can be detected by X-ray or neutron scattering only through the effects of the CDW order on the local atomic structure. In only a few materials is this effect large enough to allow for the detection of the CDW-induced peaks by X-ray or neutron scattering. However, CDW order can be detected on clean surfaces in a global manner by angle-resolved photoemission spectroscopy, ARPES, or, in the case of static CDW order, in a local probe manner by scanning tunneling microscopy, STM.

Let us now return to our discussion of stripe formation in LSCO. In the range $0.0 < q < 0.02$, LSCO is an antiferromagnetic insulator, and thus we have no clusters because the material is an insulator with no or little charge hopping, and we also have no charge stripes because we still have long-range antiferromagnetic order. For $0.02 < q < 0.05$, LSCO is an insulating spin-glass at low T, and thus we would still not expect to see clusters. However, because we no longer have long-range magnetic order, stripes composed of individual charges can and do form.[16-19] LSCO has the low-temperature-orthorhombic, LTO, structure in which a coherent tilt of the $CuO_6$ octahedra runs parallel to the orthorhombic $a_o$-axis, that is, at 45° to the tetragonal $a_t$ and $b_t$ axes, where $a_t \approx b_t$.[26] The coherent octahedral tilt provides a structural pinning mechanism along this particular axis for the dopant holes, and thus the charge stripes align in the diagonal orientation along the $a_o$-axis. At $q \approx 0.02$ the stripes are insulating with a charge density of $\approx 1$ hole/Cu, in agreement with the stripe models. In LSCO, the diagonal charge stripes are not detectable with X-ray or neutron diffraction techniques because the modulation of the local atomic structure from the charge stripe modulation is too subtle. However, the associated SDW or magnetic order is detectable in neutron scattering as incommensurate (IC) magnetic peaks. As noted above, charge stripes form antiphase domain walls with opposite $Cu^{+2}$ spins across the stripe, and thus the IC magnetic peaks have an incommensurability $\varepsilon$ that is one-half that of any CDW-induced diffraction peaks.

One can readily show that if $\rho_l$ represents the charge density per tetragonal lattice unit, $a_t$, the relation between $\varepsilon$, $\rho_l$ and q is given by,

$$\varepsilon \rho_l = \tfrac{1}{2} q \qquad (5)$$

where $\varepsilon$ and $\rho_l$ are both given in reciprocal tetragonal lattice units.

Thus, according to Eqn. (5), diagonal charge stripes along the $a_o$-axis with a charge density per Cu, $\rho_{Cu} \approx 1$, and thus a $\rho_l \approx 1/\sqrt{2}$, will give rise to two IC magnetic peaks, located in reciprocal orthorhombic lattice space at $(0,k \pm \varepsilon, 0)_o$ with an incommensurability $\varepsilon \approx (1/\sqrt{2})q$, in good agreement with experiment.[19] As q is increased in LSCO above 0.02, the incommensurability of the IC magnetic peaks increases from $\varepsilon \approx (1/\sqrt{2})q$ at $q \approx 0.02$ to $\varepsilon \approx q$ at $q \approx 0.06$.[19] We thus infer, from Eqn. (2), that the charge density of the diagonal charge stripes starts to decrease from $\rho_{Cu} \approx 1$ at $q \approx 0.02$ to $\rho_{Cu} \approx 2/3$ when $q \approx 0.06$. In Fig. 2 we depict this evolution of a diagonal charge stripe. We indicate with a dashed circle the area along the stripe that would be occupied by an incipient $(Cu)_{13}$ cluster. When the stripe-charge density is $\rho_{Cu} \approx 2/3$ along the diagonal, we can see from Fig. 2 that an incipient cluster will have $\approx 2$ holes. Now $(Cu)_{13}$ clusters can begin to form once the mobility of the holes is sufficient to permit direct Cu-Cu charge hopping up to the third-nearest neighbors. This probably occurs in LSCO for $q \approx 0.04$-0.05. However, as depicted in Fig. 2(d), if $(Cu)_{13}$ clusters are aligned along the diagonal, the $d_{x^2-y^2}$ symmetry of the charge distribution within the cluster is inconsistent with the formation of charge stripes along the diagonal. We therefore conclude that diagonal ordering of clusters is not preferred. Moreover, while diagonal charge stripes can occur in an LTO crystal if there is a structural pinning mechanism oriented along the diagonal, diagonal cluster stripes cannot occur since a diagonal pinning mechanism will have the same effect on both the x and y lobes of the $d_{x^2-y^2}$ charge distribution and thus no net effect on the cluster. This will hold true no matter how strong the diagonal pinning mechanism might be.

## 5. LSCO: SUPERCONDUCTIVITY AND PARALLEL STRIPES

However, this does not mean that cluster stripes can never form in an LTO crystal like LSCO. According to the $(Cu)_{13}$-BEC model, once clusters are formed they will have two charges per cluster as long as q ≤ 2/13 because the singlet state is the lowest energy state.[24] Also, the model predicts that the doping threshold for superconductivity in LSCO is $q_0 \approx 0.055$, in agreement with experiment. Let us consider the situation for doping above this threshold. At T > $T_c$, the clusters are randomly oriented except for momentary short-range ordering between two clusters along the $a_t$ or $b_t$ axis when they are within an interaction distance apart. This momentary short-range ordering is a consequence of the hopping interaction between two $(Cu)_{13}$ clusters with $d_{x^2-y^2}$ charge symmetry. With the condensation to the superconducting state at $T_c$, the clusters begin to act coherently. Once it becomes energetically attractive for cluster ordering to occur, the momentary short-range cluster ordering parallel to the $a_t$ or $b_t$ axis in the normal state becomes a more permanent long-range cluster ordering along the $a_t$ or $b_t$ axis in the superconducting state. Thus parallel cluster stripes can occur in LSCO but only in the superconducting state and the onset temperature for cluster ordering $T_{cl} < T_c$, although it may only be slightly less. This appears to be in agreement with experiments on LSCO provided we assume that the onset temperature for magnetic ordering $T_m \approx T_{cl}$.[27] The $(Cu)_{13}$-BEC model thus predicts that superconductivity can promote parallel cluster ordering in an LTO crystal such as LSCO. Experiments show that both diagonal and parallel stripes are present in LSCO in the doping range 0.055 < q < 0.07.[19] This observation may be explained by the persistence of some insulating spin-glass regions up to q ≈ 0.07. In the insulating spin-glass regions, clusters are still not formed and the diagonal stripes present are charge stripes. The parallel cluster stripes are present only in the superconducting regions.

It is interesting to note that incommensurate SDW fluctuations are observed in LSCO at temperatures well above $T_c$, with correlations lengths that vary from ≈ 20 A at T ≈ 50 K to ≈ 10 A at T ≈ 300 K.[28] These fluctuations may well represent the momentary short-range parallel ordering between two interacting $(Cu)_{13}$ clusters in the normal state. We would expect that the correlation length (a radial dimension) would vary from a range of $4a_t - 5.5a_t$, or ≈15 – 20 A, at relatively low temperatures, where pair interactions are dominant, to a correlation length that → $2a_t$, or ≈ 8 A, at high temperatures where pair interactions are much more transitory due to the higher kinetic energy of the clusters with the result that we are now dealing primarily with individual clusters rather than cluster pairs.

Eqn (5) indicates that the incommensurabilities of diagonal and parallel stripes for the same value of q are not necessarily identical. For example, if we assume that, at q ≈ 0.06, $\rho_{Cu} \approx 2/3$ along the diagonal and thus that there are ≈ 2 holes per incipient cluster along the diagonal, and, of course, exactly two holes in an actual cluster in a parallel cluster stripe, we have $\rho_l(dia) = 2/(3\sqrt{2})$ while $\rho_l(par) = 1/2$. From Eqn (5) we then find that $\varepsilon(dia)/\varepsilon(par) = 1.06$. Fujita et al found that, at q = 0.06, $\varepsilon(dia)/\varepsilon(par) = 1.08 \pm 0.05$.[19] Although not necessarily identical, the two incommensurabilities are very close, and in the region where both diagonal and parallel stripes are present, $\varepsilon(dia) \approx \varepsilon(par) \approx q$.

Examples of parallel cluster stripes are shown in Fig. 3. These stripes are in the vertical direction, that is, parallel to the y or $b_t$ axis. Of course, there is an equal probability of horizontal stripes that are parallel to the x or $a_t$ axis since superconductivity-promoted parallel cluster stripes can form along either axis. We can readily see that the clusters will tend to order with direct contact between them along the stripe direction so as to maintain the same cluster symmetry along the stripe. This results in the sharing of a common outer Cu site between two clusters along the stripe axis. Sharing of a Cu site between two clusters is allowed as long as the clusters have the same symmetry and the Cu site plays the same molecular orbital hybridizing role in each cluster. These criteria are fulfilled for the cluster ordering shown in Fig. 3. Let us now discuss a number of important characteristics of these cluster stripes.

1). Sharing of the outer Cu sites along the stripe direction results in an overlap of the charge wavefunction along this direction. This in turn means that, whereas charge stripes are usually insulating, cluster stripes are always metallic. The metallic nature of the parallel stripes is a major characteristic of the cuprates.

2). Parallel cluster stripes can form along either the $a_t$ or $b_t$ axis. Thus two orthogonal sets of stripes are possible, one vertical and the other horizontal. This CDW order, if detectable by X-ray or neutron diffraction, would give rise to four CDW diffraction peaks located at $(\pm 2\varepsilon, k, l)_t$ and $(h, \pm 2\varepsilon, l)_t$ in tetragonal reciprocal lattice space. The coupled SDW order results in four IC magnetic peaks in a neutron scattering spectrum at $(\frac{1}{2} \pm \varepsilon, \frac{1}{2}, 0)_t$ and $(\frac{1}{2}, \frac{1}{2} \pm \varepsilon, 0)_t$. As long as q $\leq 2/13$, each cluster has two holes, and thus the average charge density along a parallel cluster stripe is fixed at $\rho_{Cu} \approx 0.5$ holes/Cu. For a parallel stripe with a fixed charge density of $\rho_{Cu} \approx 0.5$ holes/Cu, $\rho_l = 1/2$, and from Eqn. (5) we can see that the incommensurability is then fixed at $\varepsilon \approx q$. Thus parallel cluster stripes account in a natural way for a charge density of $\approx 0.5$ holes/Cu along the stripe direction and for the well known but surprising incommensurability relation $\varepsilon \approx q$ over a wide doping range.

3). In Fig. 3 we depict the cluster stripes for q = 1/14, 1/12, 1/11 and 1/10. The q = 1/14, 1/12 and 1/10-patterns are ordinal configurations since the centerlines of adjacent cluster stripes are separated by an integer multiple of $a_t$. This separation is $7a_t$, $6a_t$ and $5a_t$ for the q = 1/14, 1/12 and 1/10-configurations respectively. Note that the coupled SDW patterns would then have separations of $14a_t$, $12a_t$ and $10a_t$ for the q = 1/14, 1/12 and 1/10-configurations, in keeping with the $\varepsilon \approx q$ relation. From Fig. 3 we also see that the minimum separations between clusters in adjacent stripes are $3a_t$, $2a_t$ and one $a_t$ for the 1/14, 1/12 and 1/10-configurations respectively. Since the separations can only change by an integer value of $a_t$, a configuration intermediate between two ordinal configurations must be composed of somewhat disordered cluster stripes wherein the separation varies appropriately between the two neighboring ordinal values along the length of the stripes. This is depicted in Fig. 3 for the 1/11-configuration, where half of the length of the stripes has a 1/12-separation and the other half has a 1/10-separation. The disorder, or glassy-type order, of intermediate configurations, compared to the ordinal configurations, should be detectable by studying the CDW or SDW diffraction peaks. Tranquada has done this for a q = 0.12 stripe configuration in LNSCO.[29] His analysis indicates a glassy-type of pattern for the SDW order with 75% of the length having a separation of $8a_t$, the 1/8-configuration, and 25% of the length with a separation of $10a_t$, the 1/10-configuration. Our analysis suggests that a q = 0.12 configuration would be composed of 80% 1/8-configuration and 20% 1/10-configuration, in good agreement with Tranquada's analysis.

4). In the $CuO_2$ plane, a $d_{x^2-y^2}$ cluster charge symmetry changes to a $d_{-(x^2-y^2)}$ symmetry as the cluster is moved one lattice unit along either the $a_t$ or $b_t$ axis. Thus, we can see from Fig. 3 that for the ordinal 1/12-configuration, which has the same phase for all the stripes, there are as many as four possible cluster stripe configurations, two vertical with opposite phase and two horizontal with opposite phase. However, the opposite phase cluster stripes will produce identical SDW patterns and thus are not distinguishable in the IC magnetic peaks, although the vertical and horizontal cluster stripes are distinguishable since they produce vertical and horizontal SDW orders respectively. Thus, even though four different configurations are theoretically possible for the 1/12-configuration, experimentally one will distinguish only two types of cluster stripes unless one can employ a methodology that is sensitive to the phase of the charge wavefunction. The 1/14 and 1/10-configurations, on the other hand, are composed of vertical stripes that alternate in phase, and thus only two types of stripe patterns are possible, vertical and horizontal.

5). While charge stripes are one-dimensional, cluster stripes are not entirely one-dimensional. Again referring to Fig. 3, we see that, while there is a long-range one-dimensional charge distribution along the stripe direction, the $d_{x^2-y^2}$ charge symmetry in each cluster imparts a short-range charge distribution normal to the stripe direction as well. This may account for the observation that ARPES experiments on stripes in LNSCO are not consistent with a purely one-dimensional charge distribution but instead appear to indicate the presence of some charge distribution and charge motion perpendicular to the stripe axis as well.[30, 31]

6). The linewidths of the IC magnetic peaks are related to the degree of dynamic behavior of the stripes, and inversely related to the SDW correlation length. Highly dynamic stripes have short correlation lengths and result in broad inelastic IC magnetic peaks. Static and quasistatic stripes have long correlation lengths and result in sharp elastic IC magnetic peaks. Charge stripe models predict that the dynamic nature of the stripes will increase with doping as the mobility of the dopant charges increases. We expect the same to apply to cluster stripes. This prediction agrees with experiment for both diagonal and parallel stripes in LSCO up to q ≈ 0.07.[16, 19] However, for q > 0.07 the linewidths of the inelastic parallel magnetic peaks decreases with increasing q, reaching a minimum at q ≈ 1/8 where the magnetic peaks in LSCO appear as sharp quasistatic elastic peaks.[4, 19, 27] Above q ≈ 1/8, the IC magnetic peaks become broader inelastic peaks again.[4] We can understand this behavior for the parallel cluster stripes in the 0.055 < q < 1/8 range within the context of the $(Cu)_{13}$-BEC model as follows. Below q ≈ 0.07, the cluster stripes are far enough apart that the dynamic behavior of one stripe is essentially unaffected by the presence of other stripes. Thus, initially the dynamic behavior increases for q > 0.055 in accordance with stripe models. However at q = 1/14 or ≈ 0.07, the minimum distance between clusters is only $3a_t$ and it is reasonable to expect that this will begin to restrict the non-coherent transverse motion of individual stripes. The inhibition of transverse motion becomes even greater as q increases to 1/12 with its minimum separation of $2a_t$ and then 1/10 with a separation of only one $a_t$. Thus we would expect a rapid decrease in dynamic behavior and thus a rapid decrease in the linewidths of the IC magnetic peaks as q is increased from 1/14 to 1/8.

## 6. CONFIGURATIONAL PINNING AND THE 1/8-ANOMALY

In Fig. 4 we depict the 1/8-configuration for LSCO. We can see that this is a unique configuration for cluster ordering, because it is here that the minimum separation between the clusters in adjacent stripes goes to zero. The clusters are now in their ultimate close-packed rectilinear configuration and form a two-dimensional checkerboard pattern. However, as we can see in Fig. 4, the pattern for LSCO has an anisotropy in the $d_{x^2-y^2}$ charge distribution within the cluster. The reason for this is as follows.

The $(\psi_o)^2$ state has a $d_{x^2-y^2}$ charge symmetry and how resistant this state is to perturbation of this symmetry is dependent on the pairing strength, $\delta$, of the pre-formed bosonic pair. In LSCO, $\delta$ is quite low $\approx 1.7$ meV or $\approx 20$ K at q $\approx 1/8$.[24] This has several consequences. First, it results in relatively low values for $T_c$. Secondly, while the $(Cu)_{13}$-BEC model assumes only weak interactions between clusters, this applies only for clusters with a minimum separation of at least one $a_t$. In the 1/8-configuration the minimum transverse separation is zero, and thus transverse Coulomb repulsion effects can no longer be ignored. Of course, the cluster stripes themselves experience a longitudinal Coulomb repulsion effect as well. However, as discussed in Sec. 3, the gain in spin potential compensates for the longitudinal Coulomb repulsion effects along the stripe direction, and thus a one-dimensional cluster stripe formation does not, in itself, result in a perturbation of the $d_{x^2-y^2}$ charge distribution. However, the transverse Coulomb repulsion effect that comes into play in the two-dimensional checkerboard 1/8-configuration can perturb the $d_{x^2-y^2}$ charge distribution in LSCO and its related compounds because of the relative weakness of the $(\psi_o)^2$ state in these materials. This perturbation will transfer charge from the transverse to the longitudinal orbital lobes, and thus introduce an anisotropy in the $d_{x^2-y^2}$ charge distribution, i.e. an anisotropy in the local density of states, LDOS, between the vertical (longitudinal) and transverse (horizontal) rungs of the checkerboard CDW pattern. This perturbation or anisotropy in the LDOS is depicted in Fig. 4 as an anisotropy in the widths of the vertical and transverse lobes of the $d_{x^2-y^2}$ orbitals. In LSCO this perturbation of the $d_{x^2-y^2}$ charge distribution in the 1/8-configuration can be considered to be a form of configuration-induced charge pinning.

The unique 1/8-configuration and its associated configuration pinning can affect the superconductivity in LSCO. A perturbation of the $d_{x^2-y^2}$ charge symmetry in the $(Cu)_{13}$ clusters can come about through an admixture of higher-energy non-bosonic states of non-$d_{x^2-y^2}$ symmetry with the bosonic $(\psi_o)^2$ singlet ground state. Since the total charge remains constant, this admixture results in a decrease in the density of the bosonic quasiparticle states. We can incorporate this concept into the $(Cu)_{13}$-BEC model by multiplying the left-hand side of Eqn (2) by a factor f where f < 1, and f = $n_b'/n_b$, where $n_b$' and $n_b$ are the boson quasiparticle densities for the perturbed and unperturbed states respectively. One can readily see from Eqn (2) that when $f < 1$, $P(\delta,T)$ must increase correspondingly since the right-hand side remains constant. Thus, $T_c$ must decrease if $\delta$ remains unchanged. In Fig. 5, we show the experimental values of the mid-point $T_c$ for LSCO as a function of q.[22] There is a noticeable dip in the mid-point $T_c$ at q $\approx$ 1/8, the celebrated 1/8-anomaly. We also show the corresponding value of f, as calculated from the $(Cu)_{13}$-BEC model that reproduces these $T_c$ results. We see that the model predicts that f dips from its unperturbed value of 1, at q $\leq$ 0.10 to $\approx$ 0.8 at q $\approx$ 1/8 and then returns to 1 by q $\approx$ 0.14. Thus a $\approx$ 20% decrease in f results in a $\approx$ 20% decrease in $T_c$. The model may also provide an explanation for why it is the mid-point $T_c$ in LSCO, rather than the onset $T_c$, that exhibits the 1/8-anomaly. As discussed in Sec. 5, parallel cluster stripes in LSCO are promoted by the onset of superconductivity, and thus $T_{cl} < T_c$ although only slightly so in the case of LSCO. Thus the onset of superconductivity must precede the onset of cluster ordering with the result that the onset $T_c$ will not be affected by the configuration pinning effects of the 1/8-configuration.

The 1/8-configuration forms a two-dimensional checkerboard CDW order, and in the absence of any perturbation of the $d_{x^2-y^2}$ cluster charge distribution, that is, in the case of an isotropic LDOS, one would expect the coupled SDW order to be diagonal.[12] Since the SDW unit cell must always be twice that of the CDW unit cell in the cuprates, that is $8a_t$ at q = 1/8, a checkerboard CDW configuration oriented along the $a_t$ and $b_t$ axes can result in only a diagonal or parallel SDW configuration. The orientation of the SDW order can be expected to depend on the degree of anisotropy of the LDOS between the vertical and horizontal rungs of the checkerboard CDW pattern. Little or no anisotropy results in an essentially isotropic two-dimensional checkerboard CDW pattern and this should produce a diagonal SDW order, while a moderate or severe anisotropy results in an anisotropic checkerboard CDW pattern that is quasi-one-dimensional and this should favor a parallel SDW order. The experimental results on LSCO[4] indicate that the moderate anisotropy arising from the 1/8-configuration pinning is sufficient to result in a parallel 1/8-SDW order oriented along the direction of the greater LDOS.

For q > 1/8, there are three different possibilities: i) new non-rectilinear ( i.e. not parallel to the $a_t$ and $b_t$ axes) cluster arrangements representing higher q values can begin to appear; ii) the 1/8-configuration can persist with the additional charge occupying higher-energy non-bosonic cluster states that do not have $d_{x^2-y^2}$ symmetry; iii) the 1/8-configuration can persist with the additional charge occupying Cu sites in the spin domains. Energy considerations will determine which combination of these possibilities will come into play. Persistence of the close-packed 1/8-configuration for q > 1/8 can be expected to depend on the relative strength of any pinning mechanism along a parallel $a_t$ or $b_t$ direction. Interestingly, Yamada *et al*[4] observed that $\varepsilon \approx$ 1/8 even for q > 0.2 indicating continued persistence of some 1/8-configuration to fairly high doping levels in spite of only a moderate 1/8-configuration pinning effect. We can now add some additional elements to the list of properties of the cluster stripes.

7). As we noted above, the fact that each $(Cu)_{13}$ cluster has two holes and that the clusters will order only in the parallel orientation results in the well-known linear relation between the incommensurabilty of the IC magnetic peaks and the hole doping, i.e. $\varepsilon \approx q$. However, this relationship will hold only as long as the cluster stripes can continue to move closer together with increased doping. As we can see from Fig. 4, the cluster stripes are in their ultimate close-packed arrangement when $q = 1/8$. As $q$ is increased above $1/8$, the parallel stripes cannot move any closer together and thus we have a natural explanation for the observation that the incommensurability saturates at $\varepsilon \approx 1/8$ for $q \geq 1/8$. We show in Fig. 6 the measured variation of $\varepsilon$ with $q$,[4, 19] as well as the variation predicted by the $(Cu)_{13}$-BEC model. According to the model, for $0.02 < q < 0.055$ in LSCO, there can be no cluster stripes and any stripes present must be diagonal insulating charge stripes with an incommensurability $\varepsilon$ that varies in a continuous fashion from $\varepsilon \approx 0.014$ at $q \approx 0.02$ ($\varepsilon \approx (1/\sqrt{2})q$) to $\varepsilon \approx 0.06$ ($\varepsilon \approx q$) at $q \approx 0.06$. For $0.055 < q < 1/8$, the model predicts parallel metallic cluster stripes with an incommensurability given by $\varepsilon = q$. At $q \geq 1/8$ the theoretical incommensurabilty saturates at $\varepsilon = 1/8$. The agreement with experiment is very good over the entire doping range.

8). The $1/8$-anomaly is the result of the distortion of the relatively weak $d_{x^2-y^2}$ cluster charge distribution in LSCO-type compounds that arises from the transverse Coulomb repulsion effect that occurs when the clusters are arranged in the close-packed $1/8$-configuration. According to the $(Cu)_{13}$-BEC model this distortion results in a decrease in the density of the bosonic quasiparticles, and thus the superfluid density, and hence to a decrease in $T_c$. This then accounts for the celebrated $1/8$-anomaly in LSCO-type compounds.

9). The $1/8$-configuration is a unique configuration since it is the ultimate close-packed configuration possible with parallel $(Cu)_{13}$ cluster stripes. This configuration maximally inhibits transverse cluster motion and thus represents the most dynamically inhibited configuration for LSCO. Furthermore the distortion of the charge distribution acts as a form of charge pinning along the stripe direction. In principle, the two-dimensional checkerboard pattern of the $1/8$-configuration permits another form of stripe dynamic, oscillations between vertical and horizontal stripe directions. In LSCO, such oscillations are greatly inhibited by the configuration pinning and the attendant LDOS anisotropy. The inhibition of both the transverse motion of clusters and the oscillations between vertical and horizontal stripe directions in LSCO at $q \approx 1/8$ results in a static or quasistatic nature of the CDW and SDW orders and explains why the IC magnetic peaks in LSCO at $q \approx 1/8$ appear as sharp elastic peaks.[27]

## 7. THE EFFECT OF PINNING ON LDOS ANISOTROPY, SUPERFLUID DENSITY AND PSUEDOGAP

Pinning perturbs the $d_{x^2-y^2}$ symmetry of the $(\psi_o)^2$ bosonic cluster state. As we noted above, this results in a decrease in the density of bosonic quasiparticles due to the admixture of non-bosonic states into the $(\psi_o)^2$ bosonic ground state. Since $f = n_b'/n_b$, then $f$ must also $= n_{sc}'/n_{sc}$ where $n_{sc}'$ and $n_{sc}$ are the superfluid densities of the perturbed and unperturbed cluster states. Fig. 5 then indicates that the superfluid density in LSCO should dip by $\approx 20\%$ at $q \approx 1/8$, another aspect of the $1/8$-anomaly. Indeed a dip of $25 - 40\ \%$ has been observed in the superfluid density of LSCO at $q \approx 1/8$ in μSR experiments.[32, 33] The perturbation factor, f, also is a measure of the anisotropy of the $d_{x^2-y^2}$ charge distribution, that is, of the anisotropy in the LDOS of the charge distribution. The anisotropy will be proportional to 1/f, since anisotropy is usually represented by a quantity > 1. Therefore, the $(Cu)_{13}$-BEC model predicts that there is a direct inverse correlation between the superfluid density and the LDOS anisotropy of the cluster bosonic state, with the superfluid density decreasing as the anisotropy increases. In addition, pinning and its associated perturbation of the cluster charge distribution, will have an effect on the psuedogap of the cuprate material. The non-bosonic states that are admixed with the bosonic ground state are, of course, normal states at all temperatures, and thus the reduction in the superfluid density is accompanied by a corresponding increase in the normal state density. That is, the perturbation of the cluster charge distribution results in a shift of states from superconducting (above the superconducting gap in energy) to normal (below the superconducting gap) and thus into the psudogap. We would therefore expect that as the LDOS anisotropy increases, the superfluid density decreases and so does the magnitude of the psuedogap as the mid-gap states become occupied.

## 8. LNSCO: DIAGONAL STRIPES

The main difference between LSCO and LNSCO is that whereas LSCO has the LTO structure throughout the entire doping range, LNSCO has a low-temperature-less-orthorhombic, LTLO, or Pccn, phase at low q, which then transforms to a low-temperature-tetragonal, LTT, phase at higher values of q.[34, 35] For a Nd doping of 0.4, the LTT phase transition occurs at q ≈ 0.12, while for higher Nd doping, the LTT transition occurs at a lower value of q.[35] In the LTT phase, the octahedral tilt axis is parallel to the Cu-O-Cu bonds, that is parallel to the $a_t$ or $b_t$ axis. In the LTLO phase, the octahedral tilt axis is rotated in the $CuO_2$ plane intermediate between the diagonal orthorhombic $a_o$ or $b_o$ axis of the LTO phase and the parallel $a_t$ or $b_t$ tetragonal axis of the LTT phase. In the LTLO phase the octahedral tilt axis rotates closer to the tetragonal $a_t$ or $b_t$ axis as the crystal comes closer to the transition to the LTT phase.[34] Furthermore, the presence of the $Nd^{+3}$ ions in LNSCO produces a greater octahedral tilt angle than in LSCO[35] resulting in a significant structural pinning mechanism for both charge and cluster stripes.

According to the $(Cu)_{13}$-BEC model, in the region where LNSCO is still a spin-glass, 0.02 <q < 0.07, we will have no clusters, and the individual charges will order as insulating charge stripes along the octahedral tilt axes. This agrees with the observation of insulating diagonal charge stripes oriented along the octahedral tilt axis, which for q ≈ 0.05 is about 15° from the diagonal since at this doping LNSCO is in the LTLO phase1.[36]

## 9. LNSCO: PARALLEL STRIPES AND STRUCTURAL PINNING

Because of the stronger charge pinning present in LNSCO in the LTLO phase compared with LSCO in the LTO phase, we would expect that charge mobility in LNSCO will tend to occur at somewhat higher values of q than in LSCO, and therefore that $(Cu)_{13}$ clusters will not begin to form until q is substantially greater than 0.05. However once these clusters begin to form, there is an immediate condensation to superconductivity since the dopant level is greater than the superconductivity threshold level of 0.055. As we discussed above, diagonal cluster ordering cannot occur even with the strong structural pinning mechanism present. However, by $q \approx 0.07 - 0.08$, the octahedral tilt axis in the LTLO phase of the LNSCO crystal has rotated closer to a parallel position, and since there is now an uncompensated component of the structural pinning potential along the nearby $a_t$ or $b_t$ axis, clusters will order in the preferred parallel orientation along the tetragonal axis closest to the octahedral tilt axis. However, these parallel cluster stripes will probably be somewhat disordered with frequent transverse jogs to attempt to accommodate the direction of the local octahedral tilt axis. We depict this sort of arrangement in Fig. 7 where a $q = 1/12$ dopant concentration produces a 1/12-configuration of stripes with transverse jogs. The situation becomes much cleaner after the transition to the LTT phase, for now the octahedral tilt axis is oriented directly along the $a_t$ or $b_t$ axis and thus the clusters will order in simple parallel stripes with no transverse jogs.

In both the LTT and LTLO phases of LNSCO, parallel cluster stripes can form due to the structural pinning mechanism arising from a coherent octahedral tilt oriented not along the diagonal as in the LTO crystal structure, but either directly along one of the tetragonal axes (LTT phase) or at an oblique angle of less than 45° to one of the tetragonal axes (LTLO phase). Thus, unlike the situation in LSCO, there is no need for superconductivity to promote parallel cluster ordering in LNSCO since there is now a non-diagonal structural pinning mechanism. In addition the presence of Nd $^{+3}$ ions in LNSCO substantially increases the magnitude of the octahedral tilt angle. This results in greater pinning and thus in a higher value of $T_{cl}$,[3, 21] with $T_{cl} > T_c$ and usually significantly so because of the very low $T_c$'s in LNSCO. This is especially true in the LTT phase of LNSCO where the octahedral tilt angle is considerably greater than it is in the LTO phase of LSCO. Most importantly, the large octahedral tilt angle, combined with the fact that the tilt axis is rotated away from the diagonal towards a tetragonal axis in the LTLO phase, can result in a significant distortion or anisotropy in the $d_{x^2-y^2}$ charge distribution, i.e. in the LDOS, of the $(\psi_o)^2$ state in the $(Cu)_{13}$ clusters since the pinning effects on the x and y-lobes of the $d_{x^2-y^2}$ charge distribution are now different. Note that if the octahedral tilt axis were still oriented along the diagonal there would be no effect on the cluster charge distribution even for the large octahedral tilt angle. However, the rotation of the tilt axis away from the diagonal results in an anisotropy in the LDOS within the cluster which increases, perhaps as $(\cos\theta - 1/\sqrt{2})$, where $\theta$ is the angle between the octahedral tilt axis and the nearest tetragonal axis. Note that in Fig. 7 we have again indicated the anisotropy in the cluster $d_{x^2-y^2}$ charge distribution, i.e. in the LDOS, that arises from the structural pinning, by the relative widths of the orbital lobes. The anisotropy due to the structural pinning reaches a maximum in the LTT phase where the octahedral tilt axis is parallel to the $a_t$ or $b_t$ axis. The substantial distortion of the $d_{x^2-y^2}$ charge distribution in LNSCO in both the LTLO and LTT phases will result in a significant decrease in the perturbation factor, f, and thus in the superfluid density, and therefore in a severe suppression of superconductivity.

## 10. LNSCO: THE 1/8-CONFIGURATION: COMBINATION OF STRUCTURAL AND CONFIGURATIONAL PINNING

The distortion or anisotropy in the $d_{x^2-y^2}$ charge distribution, or in the LDOS, in LNSCO reaches its maximum at $q \approx 1/8$ where the 1/8-configurational pinning combines with the stronger LTT structural pinning. The 1/8-configuration for LNSCO will be similar to that of LSCO (Fig. 4) but the anisotropy in the LDOS will be considerably greater. In fact the combination of high structural charge pinning in the LTT phase with the added configurational pinning of the 1/8-configuration, results in an almost complete suppression of superconductivity at $q \approx 1/8$. Once the added configurational pinning weakens above $q \approx 1/8$, the superconductivity begins to reappear although still at greatly reduced $T_c$'s. In Fig. 8 we show the experimental values of $T_c$ vs q for   LNSCO.[2, 37] and the value of f calculated from the $(Cu)_{13}$-BEC model that reproduces the $T_c$ data. We see that f is below 1 throughout the entire doping range, reaching a low of $\approx 0.44$ at $q \approx 1/8$. Thus a $\approx 55\%$ drop in f is sufficient to almost totally suppress superconductivity in an LSCO-type compound.

The significant structural pinning in LNSCO at all values of q results in an appreciable LDOS anisotropy, which, according to the values of f in Fig. 8, ranges from $\approx 1.4$ to $\approx 2.5$. This anisotropy should be present in both the one-dimensional cluster stripes for $q < 1/8$ and in the two-dimensional checkerboard pattern for $q \geq 1/8$. In addition, the low values of f indicate low values for the superfluid density. Thus we expect that LNSCO will exhibit both a greater LDOS anisotropy and a smaller superfluid density than LSCO at all values of q. Finally, LNSCO should also exhibit a smaller psuedogap than LSCO.

Unlike the situation in LSCO, the structural pinning imparted by the large octahedral tilt angle in both the LTLO and LTT phases of LNSCO greatly diminishes the dynamic behavior of the cluster stripes at all values of q. In LSCO, the IC magnetic peaks are dynamic and appear as fairly broad inelastic neutron scattering peaks for most values of q except when $q \approx 1/8$. At $q \approx 1/8$, the IC magnetic peaks in LSCO are essentially quasistatic, because of the 1/8-configurational pinning and appear as sharp elastic neutron scattering peaks. In LNSCO, the structural pinning alone is sufficient to pin the clusters at all values of q, and thus sharp elastic IC magnetic peaks should be observed at all q, and not just in the vicinity of $q \approx 1/8$.[1-3]

In Fig. 9 we depict in a qualitative manner, using the relative widths of the orbital lobes of the $d_{x^2-y^2}$ charge distribution, how we might expect the anisotropy in the LDOS to vary with the degree of pinning. We indicate how the perturbation factor, f, and the anisotropy, given by 1/f, will vary. For no or very weak pinning $f \approx 1$ and there is no anisotropy, while for moderate pinning f will decrease to $\approx 0.8$ and the anisotropy will $\approx 1.25$, and finally for very strong pinning f will $\approx 0.4$ and the anisotropy will $\approx 2.5$. Although we depict, in Fig. 9, a pinning along the $b_t$ axis and the resultant vertical stripes, a similar situation will hold for the case of pinning along the $a_t$ axis and horizontal stripes.

In LNSCO, the onset temperature for cluster ordering, $T_{cl}$ is $> T_c$ at all values of q due to the combination of large structural pinning and low values of $T_c$. Furthermore, CDW ordering appears to exist homogeneously throughout the sample[38] and each $CuO_2$ layer has only one stripe direction that alternates between vertical and horizontal in adjacent $CuO_2$ layers.[2, 3] This indicates that LNSCO should exhibit interesting charge transport properties in the normal state of the LTT phase. For $T > T_{el}$, which in LNSCO appears to be set by the structural-phase LTO → LTT transistion temperature, $T_0$ (75 −85 K ),[39] the transport properties should be similar to LSCO since the clusters are randomly oriented. For $T_c < T < T_{el}$ and $q < 1/8$, the metallic parallel cluster stripes in any given $CuO_2$ layer are all ordered either along the $a_t$ or $b_t$ axis only, and thus charge transport in each $CuO_2$ layer should be one-dimensional with no transverse component. This condition will persist up to $q = 1/8$. At $q = 1/8$, the clusters order into the two-dimensional checkerboard pattern of Fig. 4, and this permits some transverse charge transport. However, the component of transverse charge transport at $q = 1/8$ will be small given the large LDOS anisotropy from the significant LTT structural pinning. Thus at $q = 1/8$ one should expect an onset of some two-dimensional charge transport. For $q > 1/8$, the appearance of new non-rectilinear arrangements, or the addition of non-$d_{x^2-y^2}$ symmetry charge in the clusters, or the presence of additional charge within the spin domains, breaks the parallel symmetry and makes two-dimensional charge transport even more possible. Thus one-dimensional charge transport should be present for $q < 1/8$, giving way to increasing two-dimensional charge transport for $q \geq 1/8$. This unusual charge transport behavior has apparently been observed by Noda *et al* in Hall coefficient experiments on an LNSCO compound with 0.6 Nd.[39]

## 11. LBSCO

LBSCO is an LSCO-type compound of particular interest. When the total doping of (Ba + Sr) ions < 0.12, the compound is a superconductor with a reasonable $T_c$. However, when (Ba + Sr) $\approx 1/8$, the hole doping is $\approx 1/8$ and superconductivity is suppressed, another example of the 1/8-anomaly.[23, 40] Of particular interest is the fact that the crystal structure of LBSCO$_{q = 1/8}$ ( fixed hole doping of 1/8) varies with Sr doping, starting with an LTT phase at low Sr doping, then going to a LTLO phase and finally to an LTO phase at higher Sr doping.[40] Experimentally one finds that, in both the LTT and LTLO phases, the superconductivity of LBSCO$_{q = 1/8}$ is severely suppressed, whereas in the LTO phase the superconductivity is only mildly suppressed. Furthermore, while the robustness of the CDW order is appreciable in the LTT and LTLO phases, it is greatly reduced in the LTO phase. On the other hand the SDW order exhibits static or quasistatic behavior in all phases. Finally, while the stripes observed are parallel in all phases, the stripes in the LTLO phase appear more disordered and mosaic-like than in the LTT phase.[41] All of these observations of  LBSCO$_{q = 1/8}$ may be explained by the $(Cu)_{13}$-BEC model.

In the LTT phase, one would expect that LBSCO$_{q = 1/8}$ would behave similarly to LNSCO with $q \approx 1/8$. The combination of the strong structural pinning and the weaker 1/8-configurational pinning will significantly perturb the $d_{x^2-y^2}$ charge symmetry, or LDOS, of the $(Cu)_{13}$ clusters resulting in a severe suppression of superconductivity. In the LTLO phase, the octahedral tilt axis is rotated away from the $a_t$ or $b_t$ axis, but there is still a significant component of the structural pinning potential along the nearest tetragonal axis. This condition is similar to LNSCO in its LTLO phase ($q < 0.12$), with the result that in LBSCO$_{q=1/8}$ in the LTLO phase there is still significant structural pinning along the tetragonal axis nearest to the octahedral tilt axis, and this combines with the 1/8-configurational pinning to severely suppress superconductivity. However, the extent of the $d_{x^2-y^2}$ distortion, or LDOS anisotropy, and thus the extent of the superconductivity suppression, is a function of how far the octahedral tilt axis is rotated away from the tetragonal axis, i.e. the magnitude of $(\cos\theta - 1/\sqrt{2})$. As the Sr doping is increased and LBSCO$_{q=1/8}$ approaches the LTO phase, $\theta$ increases towards 45°, and we would expect the degree of structural pinning to decrease and with it the degree of superconductivity suppression. Finally, in the LTO phase, where the octahedral tilt axis is along the diagonal, the effect of structural pinning on the (Cu)$_{13}$ cluster totally disappears. Nevertheless, since q = 1/8, we still have the 1/8-configurational pinning. Thus LBSCO$_{q=1/8}$ in the LTO phase should be similar to LSCO at q ≈ 1/8, and thus exhibit only a mild suppression of superconductivity.

In the LTT and LTLO phases of LBSCO$_{q=1/8}$, we have strong structural pinning in addition to the 1/8-configurational pinning. This results in robust parallel cluster stripes, which are detectable with neutron or X-ray diffraction techniques through their effect on the local atomic order. The accompanying SDW orders will be static and appear as sharp elastic IC magnetic peaks in neutron scattering. The cluster stripes in the LTLO phase will be more disordered than in the LTT phase since the octahedral tilt axis is now rotated away from the tetragonal axis. While the (Cu)$_{13}$ clusters will still order as parallel stripes along the Cu-O-Cu bonds, they will exhibit frequent transverse jogs to accommodate the oblique orientation of the local octahedral tilt axis (Fig. 7), thereby exhibiting more disorder. In the close-packed 1/8-configuration, this disorder will result in the observed mosaic-type of pattern of the LTLO phase. In the LTO phase, the CDW order is much weaker than in the LTLO or LTT phases, since there is no structural pinning potential and there is only the relatively weak 1/8-configurational pinning. The effects on the local atomic structure are considerably reduced, and as in LSCO, the CDW is not detectable with X-ray or neutron diffraction. Nevertheless, just as in LSCO, the CDW order is still there, and it is dynamically inhibited by the 1/8-configuration pinning effect, with the result that the accompanying SDW orders are quasistatic and exhibit sharp elastic IC magnetic peaks. In summary, the (Cu)$_{13}$ cluster concept together with the (Cu)$_{13}$-BEC model appears to fully account for the results seen in LBSCO.[40, 41]

The model further predicts that the LDOS anisotropy, the superfluid density and the psuedogap should also vary in LBSCO as the Sr concentration is changed. In the LTO phase, The LDOS anisotropy should be relatively mild, and the superfluid density and psuedogap should be similar to LSCO at q ≈ 1/8. As the compound enters the LTLO phase, the anisotropy should rapidly increase, and the superfluid density and psuedogap rapidly decrease, as the octahedral tilt axis rotates towards a tetragonal axis. Finally in the LTT phase, LBSCO should exhibit a large LDOS anisotropy and a small superfluid density and psudogap similar to what we would expect in LNSCO at q ≈ 1/8.

## 12. CUPRATES WITH HIGHER Tc's

In higher-$T_c$ cuprates such as $YBa_2Cu_3O_{7-y}$, Y123, and $Bi_2Sr_2CaCu_2O_{8-y}$, Bi2212, we would expect to see parallel cluster ordering as well. And indeed parallel SDW order has been seen in Y123, indicating the presence of parallel CDW order.[42-45] As with LSCO, Y123 has an orthorhombic crystal structure and thus, according to the model, will experience no structural pinning of the clusters, resulting in dynamic parallel CDW and SDW orders. This seems to be borne out in Y123, where the SDW order is characterized by a quartet of broad inelastic parallel IC magnetic peaks over the entire doping range.[42-45] The lack of structural pinning effects on clusters in orthorhombic crystals like LSCO and Y123 makes the parallel CDW order difficult to detect because the effect on the local atomic order is too subtle, but it also implies that the parallel CDW and SDW orders will coexist with superconductivity. However, there is an important difference between the higher-$T_c$ cuprates and the LSCO-type compounds. The binding energy of the $(\psi_o)^2$ state in Y123 and Bi2212 is 2-3X larger than in LSCO. This results not only in a higher $T_c$, but also in a more robust $(\psi_o)^2$ state and thus in a more robust $d_{x^2-y^2}$ charge distribution. We therefore do not expect that the relatively weak 1/8-configuration pinning will have much effect on the $d_{x^2-y^2}$ cluster charge distribution in Y123 or in other higher-$T_c$ cuprates. This then appears to account for the absence of a 1/8-anomaly in the higher-$T_c$ cuprates and in the presence of broad inelastic IC magnetic peaks in Y123 at $q \approx 1/8$.

A striking experimental observation in Y123 is that while the incommensurability $\varepsilon$ of the IC magnetic peaks still varies linearly with q, it does not saturate at a value of $\varepsilon \approx 1/8$ for $q \geq 1/8$, but rather saturates at a value of $\varepsilon \approx 0.10$ for $q \geq 0.10$.[45] It is possible to explain this observation as a consequence of the more robust $d_{x^2-y^2}$ charge distribution. At $q \approx 1/8$ in Y123 or Bi2212, we have the unique close-packed two-dimensional 1/8-configuration. However, unlike the situation in LSCO, the robustness of the $d_{x^2-y^2}$ charge distribution makes it much more resistant to the perturbation of the transverse Coulomb repulsion effect that comes into play at the 1/8-configuration. The result is that the 1/8-configuration will be, as depicted in Fig 10, essentially an isotropic two-dimensional checkerboard pattern with little anisotropy in the LDOS of the vertical and horizontal rungs of the pattern. As we discussed in Sec. 5, an isotropic two-dimensional checkerboard CDW order with incommensurability $2\varepsilon$ will result in a *diagonal* SDW order of incommensurability $\varepsilon$. Unfortunately, neutron scattering experiments set up to detect parallel SDW order usually cannot detect diagonal order as well, so the diagonal SDW order that accompanies the checkerboard CDW order with an isotropic LDOS may go undetected. As noted previously, the ordinal configurations are 1/14, 1/12, 1/0 and 1/8. Intermediate values of q have stripe configurations with a spacing along the stripe that alternates between those of the nearest ordinal configurations. Thus the ordinal 1/10 and 1/8 configurations combine to make up the CDW orders for 1/10 < q < 1/8. However a neutron scattering geometry set up to detect parallel SDW order in this doping range will see only the parallel IC magnetic peaks from the 1/10-configuration and not the diagonal IC magnetic peaks from the 1/8-configuration. In this situation, the 1/10-configuration will then appear as the saturation parallel configuration in Y123, and the incommensurability $\varepsilon$ for parallel SDW order will then appear to saturate at a value of $\varepsilon \approx 1/10$ starting at $q \approx 1/10$ and going to at least $q \approx 1/8$. Since the robustness of the $(\psi_o)^2$ state and its $d_{x^2-y^2}$ charge distribution is dependent on the magnitude of the preformed pair binding energy and since this binding energy decreases with doping, we can expect that at some doping level $> 1/8$, the 1/8-configuration pinning effect will begin to distort the $d_{x^2-y^2}$ charge distribution and introduce an anisotropy in the LDOS of the 1/8-CDW pattern. Based on the results in LSCO, the increasing anisotropy with doping should result in a transition from the diagonal back to the parallel SDW pattern in the slightly overdoped region when the binding energy drops to $\approx 1.7$ meV. In Y123, this will occur in the slightly overdoped region where $T_c \approx 85$ K. Thus we expect a diagonal SDW order in Y123 to begin to appear at $q > 1/10$ and then disappears for $q > 0.19$ ($\leq 85$ K on the slightly overdoped side). An important test of the $(Cu)_{13}$-BEC model would be the detection of the diagonal IC magnetic peaks with an incommensurability $\varepsilon \approx 1/8$ that characterize the expected $q \approx 1/8$ SDW order in Y123.

## 13. EXPERIMENTAL EVIDENCE OF THE 1/8-CONFIGURATION

Returning to Figs. 4 and 10, we see that the two-dimensional checkerboard pattern of the 1/8-configuration is parallel to the Cu-O-Cu bonds, that is, the rungs of the checkerboard pattern run parallel to the $a_t$ and $b_t$ axes, and that the spacing between both the vertical and horizontal rungs is $4a_t$. This particular checkerboard CDW order appears to have been imaged in STM experiments by Hoffman *et al*[46] and Howald *et al*[47] in Bi2212 for $q > 1/8$. In addition, these experiments involve static CDW orders, arising from pinning from vortices[46] and from surface point defects[47]. Since these pinning mechanisms are not confined to a diagonal axis, they will act as a local non-diagonal structural pinning mechanism and consequently perturb the $d_{x^2-y^2}$ cluster charge distribution. We therefore expect that the 1/8-configurations imaged in these experiments will exhibit an anisotropic $d_{x^2-y^2}$ charge distribution similar to that depicted in Fig. 4 rather than the isotropic configuration of Fig. 10. This should appear in an STM image as an anisotropic LDOS between the vertical and horizontal rungs of the two-dimensional checkerboard CDW patterns. This is exactly what has been observed. Hoffman *et al* have found an LDOS anisotropy between the vertical and horizontal rungs of the two-dimensional checkerboard pattern in the vicinity of a vortex of $\approx 3$,[46] while Howald *et al* saw an anisotropy of $\approx 1.6$ in the two-dimensional checkerboard pattern that is pinned by surface point defects.[47] Furthermore, Howald *et al* report that the checkerboard CDW pattern appears to be correlated with a shift of states from above to below the superconducting gap, consistent with the predictions of the model that pinning will shift states from superconducting (above the gap) to normal (below the gap). Recently, other weak two-dimensional LDOS patterns have been observed in STM images in field-free samples of Bi2212 [48, 49]. Unlike the patterns imaged in Refs *46* and *47*, these new LDOS patterns exhibit dispersion whereby the modulation period varies with bias voltage or energy and thus they have been interpreted as resulting from quasiparticle scattering interference effects. However, the checkerboard LDOS patterns of Refs *46* and *47* do not appear to exhibit this dispersion characteristic and, in addition, the magnitude of the LDOS patterns near the vortices is up to 100X greater than those attributed to quasiparticle scattering [48]. While LDOS patterns that exhibit dispersion may well be due to quasiparticle scattering, the non-dispersive checkerboard LDOS patterns are most likely from pinned static CDW stripes forming the 1/8-configuration. An interesting test of this hypothesis would be to investigate the STM images near the vortices as a function of magnetic field. If the patterns are due to pinned stripes, then the spacing in the checkerboard LDOS pattern should remain fixed at $4a_t$ but the LDOS anisotropy and the shift of states from above gap to below gap should both increase as the magnetic field increases. The STM images of non-dispersive LDOS checkerboard patterns reported in Refs. *46* and *47* appear to provide compelling evidence for the existence of the cluster stripe 1/8-configuration predicted for the case of moderate to severe pinning   (Fig. 4), and thus provide a direct validation of the $(Cu)_{13}$-BEC model.

The model further predicts that any pinning effect that is preferentially oriented along one of the tetragonal axes can result in a distortion of the cluster $d_{x^2-y^2}$ charge distribution and a suppression of superconductivity. The pinning effect can come from a number of different global and local sources, including crystal structure, such as the octahedral tilt angle, the close-packed 1/8-configuration, vortices and point defects, and can result in suppression of superconductivity that can range from mild to complete. The $(Cu)_{13}$-BEC model thus indicates that one should consider mechanisms that can distort the $d_{x^2-y^2}$ cluster charge symmetry, or introduce an anisotropy in the cluster LDOS, as potentially important when dealing with instances of superconductivity suppression. Beyond the situations described above for the LSCO-type compounds, we might include the suppression of superconductivity in Y123 in the tetragonal phase, the effects of non-magnetic substitutions such as $Zn^{+2}$, possible suppression outside the core of a vortex and the apparent lack of a superconducting state at the cleaved surfaces of most high-$T_c$ cuprates.

## 14. SUMMARY

Let us summarize the key characteristics of charge-spin order in the superconducting cuprates that the $(Cu)_{13}$-BEC model appears able to explain. The parallel stripes in the superconducting state of LSCO are the result of cluster stripe ordering promoted in the superconducting state by the long-range coherence of the $(\psi_o)^2$ cluster state with its $d_{x^2-y^2}$ symmetry. In other LSCO-type compounds that have an LTLO or LTT crystal structure, parallel cluster stripes are not promoted by the superconductivity but instead result from structural pinning effects. The metallic nature of parallel cluster stripes is a direct consequence of the overlap of the $d_{x^2-y^2}$ charge distribution along the stripe direction. Since $(Cu)_{13}$ clusters have two holes per cluster over a large doping range, the parallel cluster stripes have a fixed charge distribution of $\approx 0.5$ holes/Cu along the $a_t$ or $b_t$ axis, thereby resulting in the well-known linear relationship between the incommensurability and doping, $\varepsilon \approx q$. The unique close-packed 1/8-configuration results in the saturation of $\varepsilon \approx 1/8$ for $q \geq 1/8$. Furthermore, this unique 1/8-configuration and its associated configurational pinning provide a natural explanation for the celebrated 1/8-anomaly. The model also shows that the additional structural pinning that occurs in the LTLO and LTT phases of LNSCO and LBCO, that arises because of the rotation of the octahedral tilt axis towards the $a_t$ or $b_t$ axis, can result in an even greater suppression of superconductivity in these materials. The $(Cu)_{13}$-BEC model readily accounts for the principal observations about the CDW and SDW orders and their various effects on the superconductivity in LBSCO$_{q=1/8}$ where the crystal structure progressively changes from LTO to LTLO and then LTT as the concentration of Sr is increased. The model also appears to account for the variations with doping and crystal structure of the IC magnetic peak linewidths, and thus of the degree of dynamic behavior of the SDW orders, in the various LSCO-type compounds. The concept of (Cu)$_{13}$ cluster stripes may explain the puzzling ARPES results on LNSCO stripes which indicate the presence of some charge distribution normal to the stripe axis. Similarly, the unusual charge transport properties of LNSCO in its normal state may be accounted for by the presence of cluster stripes. The experimental results in higher-T$_c$ cuprates such as Y123 may also be explained by the (Cu)$_{13}$-BEC model which predicts a much more isotropic LDOS in the field-free bulk material. Most importantly, STM experiments on Bi2212 samples, in the presence of local pinning effects, show two-dimensional checkerboard CDW patterns with an anisotropic LDOS, that are in excellent agreement with the predictions of the model.

In addition to accounting for many of the key characteristics of the CDW and SDW orders seen in the superconducting cuprates, the (Cu)$_{13}$-BEC model also makes several other interesting predictions. Firstly, while diagonal stripes are usually insulating charge stripes, parallel cluster stripes are always metallic. Secondly, cluster stripes exhibit local $d_{x^2-y^2}$ symmetry with long-range charge distribution along the stripe axis and short-range charge distribution normal to the axis. This unusual charge distribution should show up in STM experiments on the static one-dimensional stripes in LNSCO for q < 1/8, if one can obtain samples with clean smooth surfaces. The CDW pattern in LNSCO for q ≥ 1/8 should exhibit a two-dimensional checkerboard pattern with a pronounced anisotropy in the LDOS between the vertical and horizontal rungs of the pattern. This LDOS anisotropy should also be observable in the one-dimensional parallel stripes. The STM experiments on LNSCO are probably best performed in the normal state, $T_{cl} < T < T_c$, where complications from the superconducting state are absent. The (Cu)$_{13}$-BEC model also predicts a close relationship between the LDOS anisotropy in the cluster and the superfluid density and psuedogap. In the presence of pinning, the LDOS anisotropy should increase while both the superfluid density and psuedogap should decrease. The LBSCO$_{q=1/8}$ compound appears to be an ideal system to test this prediction since the degree of pinning can be readily controlled through the concentration of Sr. The model also predicts that in higher-T$_c$ cuprates without pinning effects, such as bulk Y123 and Bi2212 in the absence of a magnetic field, the unique 1/8-configuration will result in a two-dimensional checkerboard CDW order with an essentially isotropic LDOS that will have an associated diagonal SDW order. Experiments to investigate these various predictions would provide good tests of the (Cu)$_{13}$-BEC theory.

In conclusion, the (Cu)$_{13}$-BEC model appears able to account for many of the key characteristics of the coupled CDW and SDW orders seen in the superconducting cuprates. This ability to account for the key magnetic properties, when combined with the previously discussed ability to account for the principal thermodynamic and electronic properties of these compounds, provides considerable additional support for the (Cu)$_{13}$-BEC model of high-temperature superconductivity.

## ACKNOWLEDGEMENTS

I thank R. Birgeneau, A. Kapitulnik, J. Tranquada and R. Laughlin for helpful discussions.

<u>REFERENCES</u>

# FIGURE CAPTIONS

Fig.1.  A depiction of two $(Cu)_{13}$ clusters, one with a $d_{x^2-y^2}$ and the other with a $d_{-(x^2-y^2)}$ charge distribution for the bosonic $(\psi_0)^2$ state of the pre-formed pair. The dots indicate the Cu lattice sites in the $CuO_2$ plane. The nominal cluster size is $4a_t$ where $a_t$ is the tetragonal lattice constant, and $b_t \approx a_t$. The unshaded and shaded lobes of the cluster $(\psi_0)^2$ wavefunction represent positive and negative phase orbitals respectively.

Fig. 2.  Evolution of diagonal charge stripes with doping: a) $q \approx 0.02$, charge density per Cu along the diagonal, $\rho_{Cu} \approx 1$, charge density per tetragonal lattice unit, $a_t$, $\rho_l \approx 1/\sqrt{2}$; b) $q \approx 0.04$, $\rho_{Cu} \approx 3/4$, $\rho_l \approx 3/(4\sqrt{2})$; c) $q \approx 0.06$, $\rho_{Cu} \approx 2/3$, $\rho_l \approx 2/(3\sqrt{2})$; d) diagonal arrangement of $(Cu)_{13}$ clusters showing incompatibility with diagonal charge ordering because of $d_{x^2-y^2}$ cluster charge symmetry. Incipient clusters are indicated in dotted outline.

Fig. 3.  Depiction, for LSCO, of parallel cluster stripes along the $b_t$ axis for the ordinal configurations at $q = 1/14$, $1/12$ and $1/10$, and for the intermediate configuration at $q = 1/11$.

Fig. 4. The close-packed 1/8-configuration at $q = 1/8$ for LSCO. The two-dimensional checkerboard pattern exhibits an anisotropy in the LDOS between the vertical and horizontal orbital lobes of the $d_{x^2-y^2}$ cluster charge distribution arising from configuration-pinning effects This anisotropy is depicted as an anisotropy in orbital lobe widths.

Fig. 5. The mid-point $T_c$ vs q for LSCO (from Ref. 22). The perturbation factor, f, ($- - - - -$) is calculated from the $(Cu)_{13}$-BEC model to reproduce the $T_c$ data. The f for an unperturbed cluster charge distribution is 1.

Fig. 6.  IC magnetic peak incommensurability $\varepsilon$ versus q for LSCO. The data is from Refs. 4 and 19, and the theoretical curve ($- - - - -$) is derived from the $(Cu)_{13}$-BEC model.  Region A exhibits insulating diagonal charge stripes, while region B exhibits metallic parallel cluster stripes. The two regions coexist for $0.055 < q < 0.07$.

Fig. 7.  Depiction of $q = 1/12$ parallel stripes for LNSCO in the LTLO phase with the octahedral tilt axis at angle $\theta$ to the $b_t$ axis. The transverse jogs attempt to accommodate the oblique angle of the local octahedral tilt axis which defines the structural pinning direction. The anisotropy in the LDOS of the clusters arising from structural pinning effects is depicted as an anisotropy in orbital widths.

Fig. 8.  $T_c$ vs q for LNSCO. (from Refs. 2 and 37). The perturbation factor, f, ($- - - - -$) is calculated from the $(Cu)_{13}$-BEC model to reproduce the $T_c$ data. The f for an unperturbed cluster is 1.

Fig. 9.  Qualitative depiction of the change in anisotropy in the cluster $d_{x^2-y^2}$ charge distribution, or LDOS, as the degree of pinning increases. The anisotropy is given by 1/f.

Fig. 10. The predicted 1/8-configuration for the higher-$T_c$ cuprates, Y123 and Bi2212, which in a field-free bulk material will form a two-dimensional checkerboard pattern with isotropic LDOS because of the absence of pinning effects. The associated SDW order will be diagonal with an incommensurability $\varepsilon \approx 1/8$.

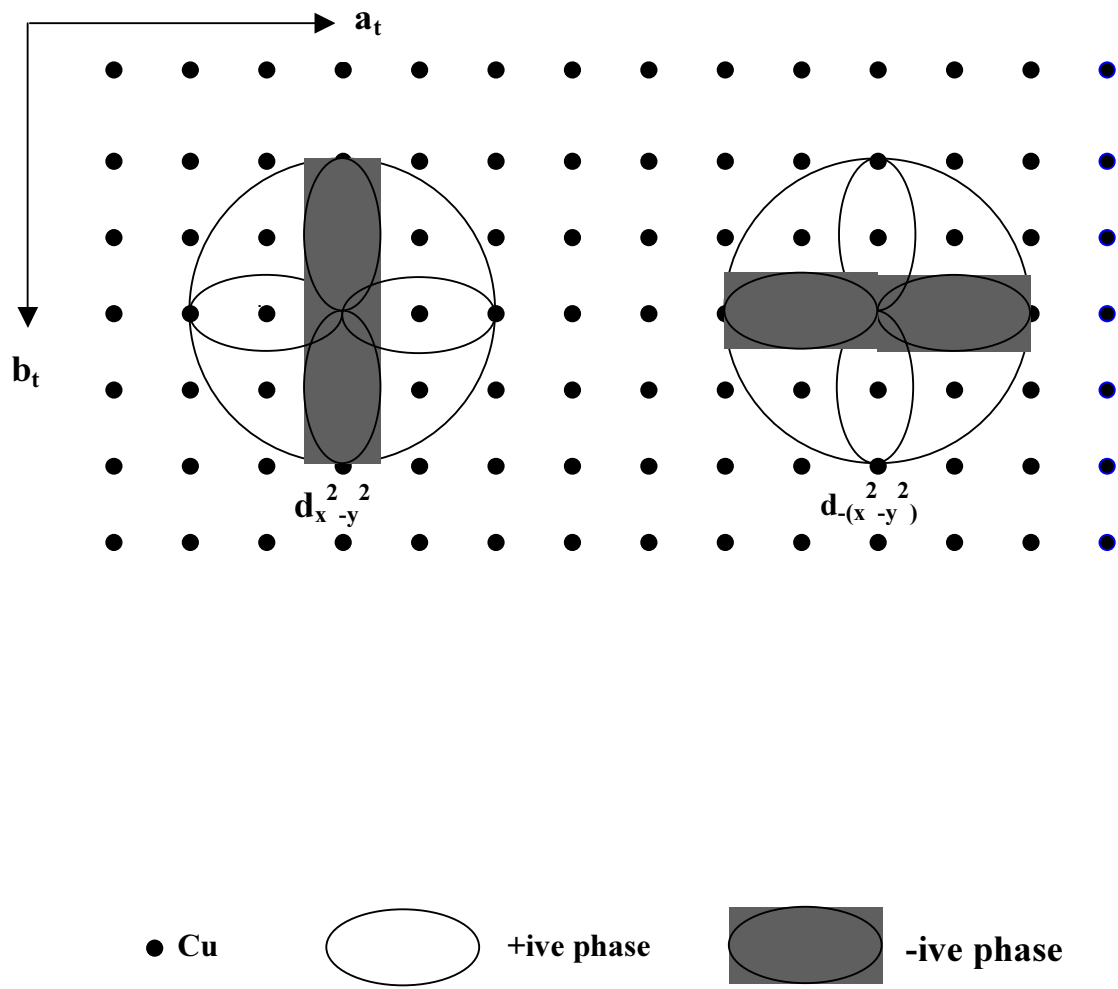

Fig. 1.  A. Rosencwaig

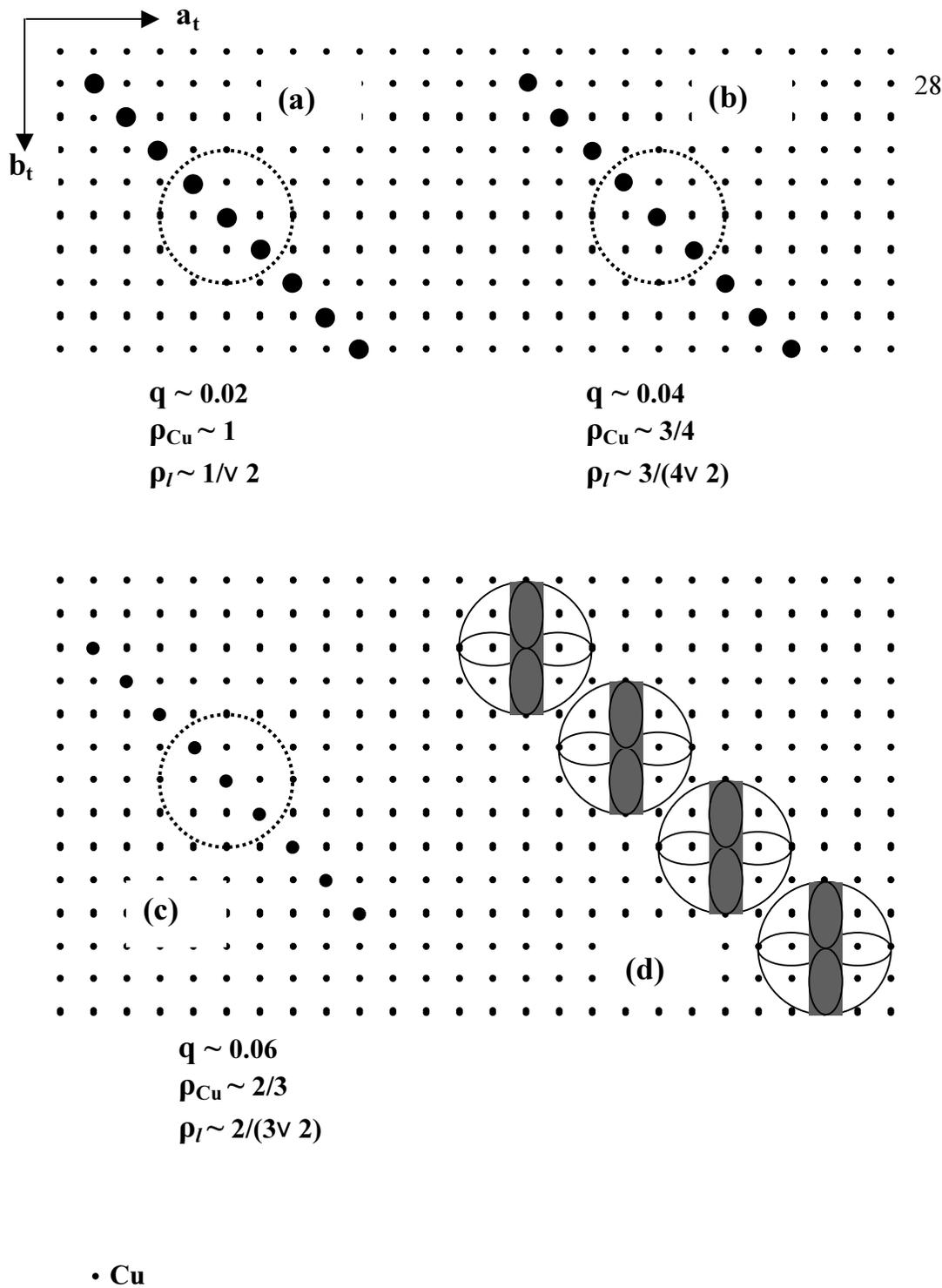

**a_t**

**b_t**

**(a)**

**(b)**

q ~ 0.02
ρ_Cu ~ 1
ρ_l ~ 1/√2

q ~ 0.04
ρ_Cu ~ 3/4
ρ_l ~ 3/(4√2)

**(c)**

**(d)**

q ~ 0.06
ρ_Cu ~ 2/3
ρ_l ~ 2/(3√2)

• **Cu**

Fig. 2. A. Rosencwaig

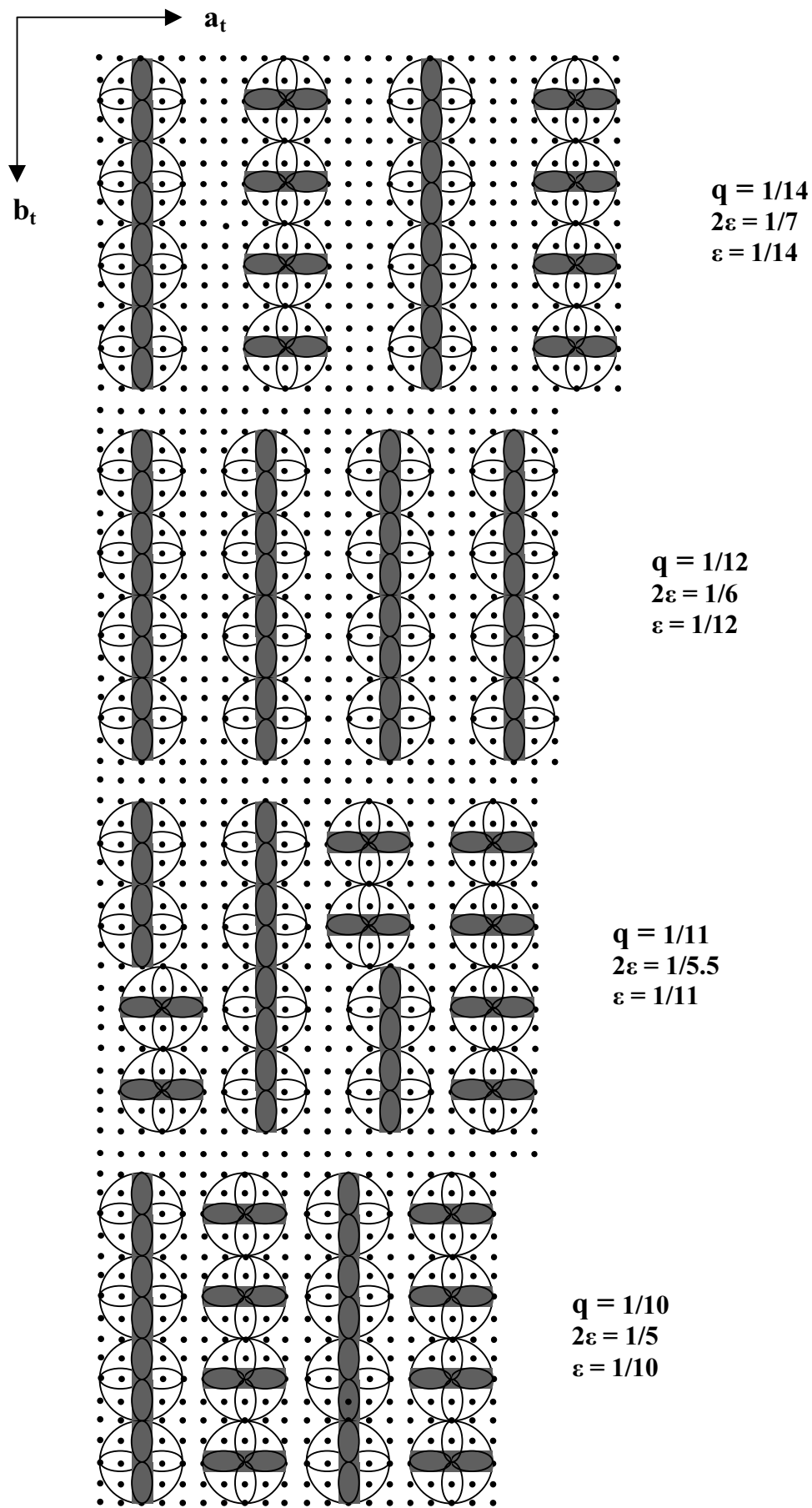

$a_t$

$b_t$

q = 1/14
2ε = 1/7
ε = 1/14

q = 1/12
2ε = 1/6
ε = 1/12

q = 1/11
2ε = 1/5.5
ε = 1/11

q = 1/10
2ε = 1/5
ε = 1/10

Fig. 3   A. Rosencwaig

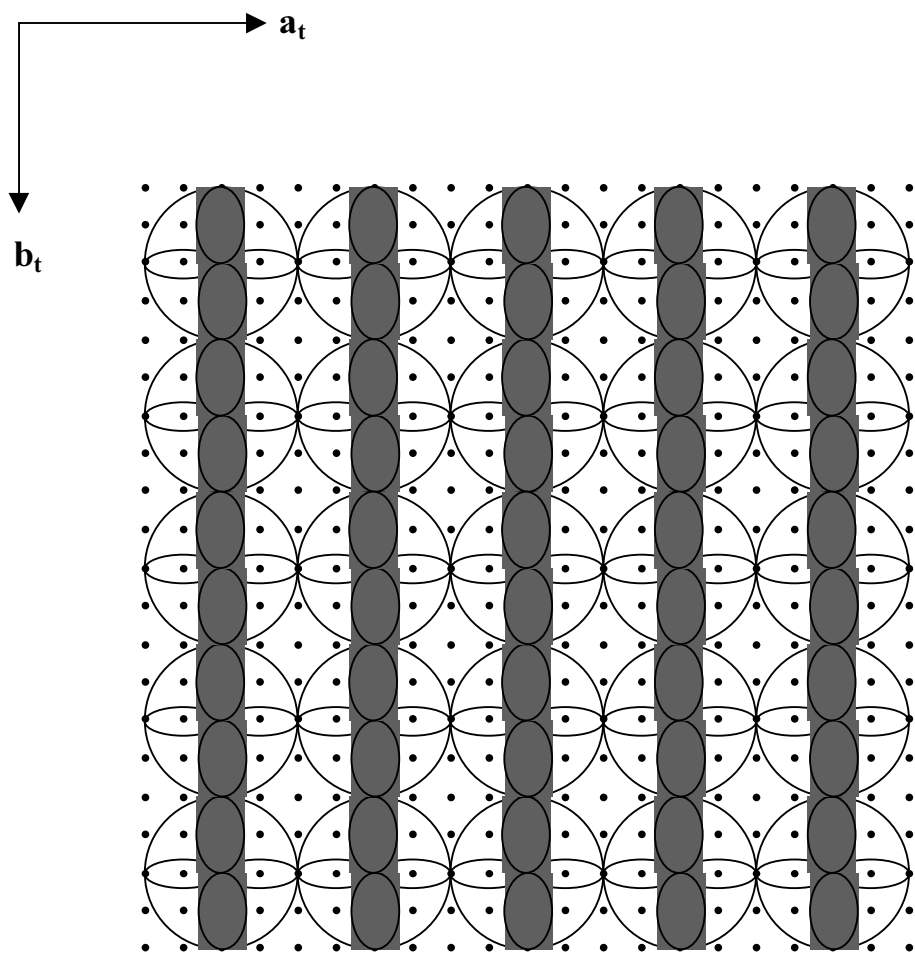

**a_t**

**b_t**

**q = 1/8**
**2ε = 1/4**
**ε = 1/8**

Fig. 4  A. Rosencwaig

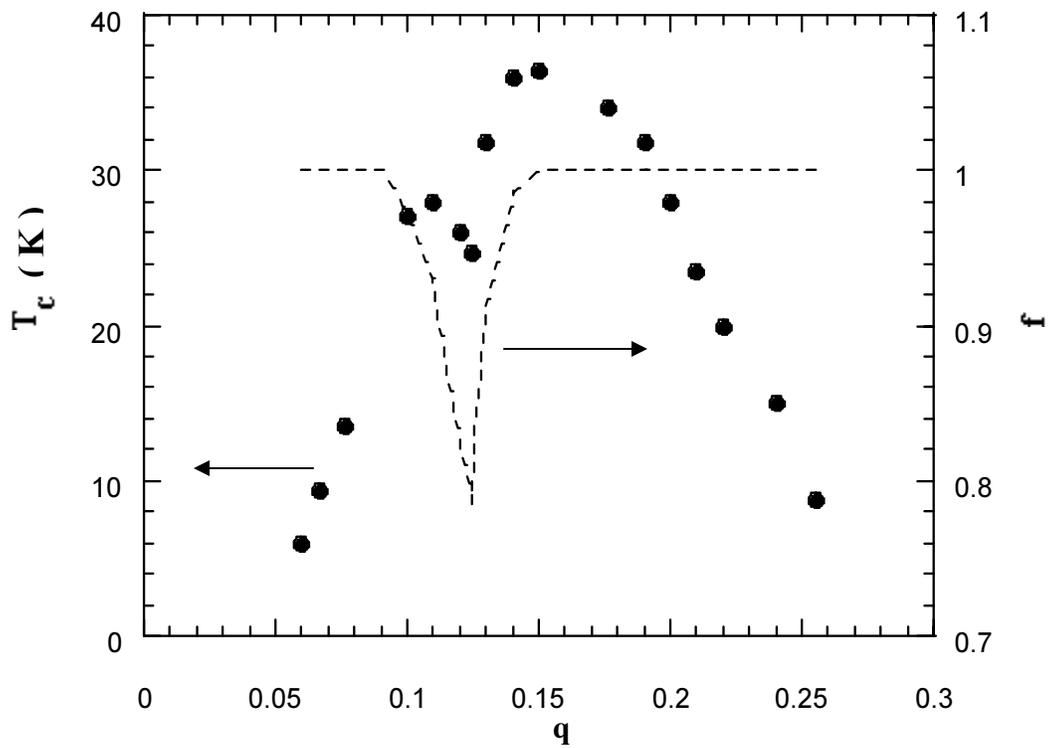

Fig. 5  A. Rosencwaig

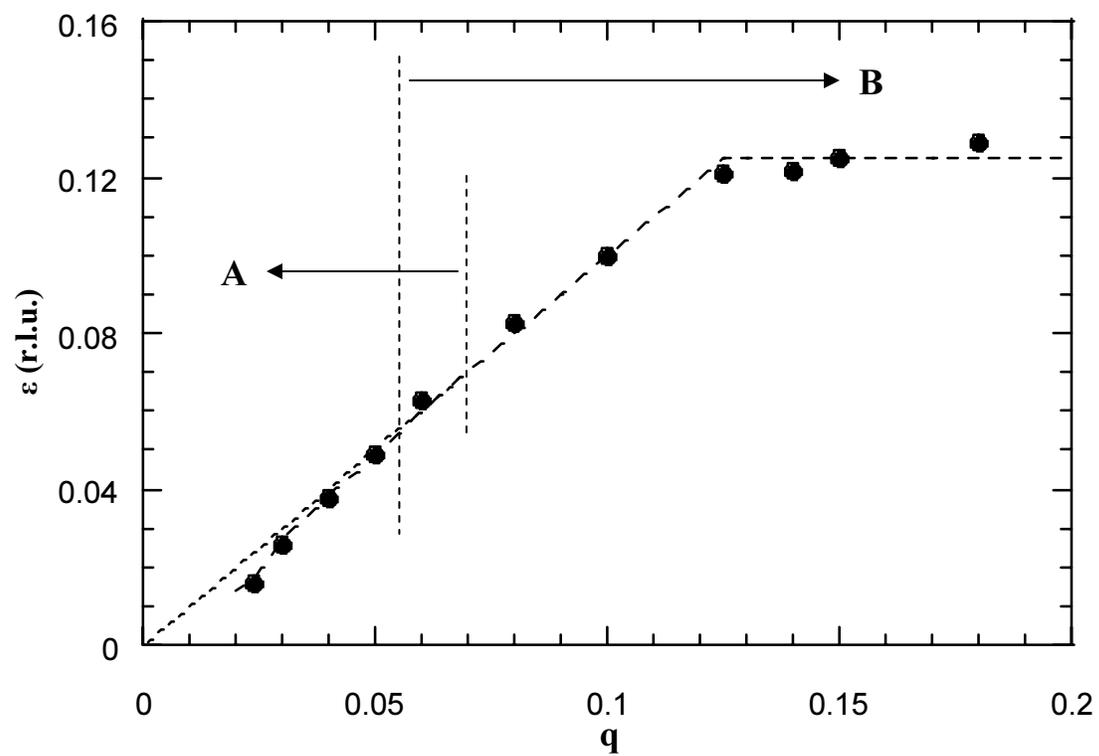

Fig. 6  A. Rosencwaig

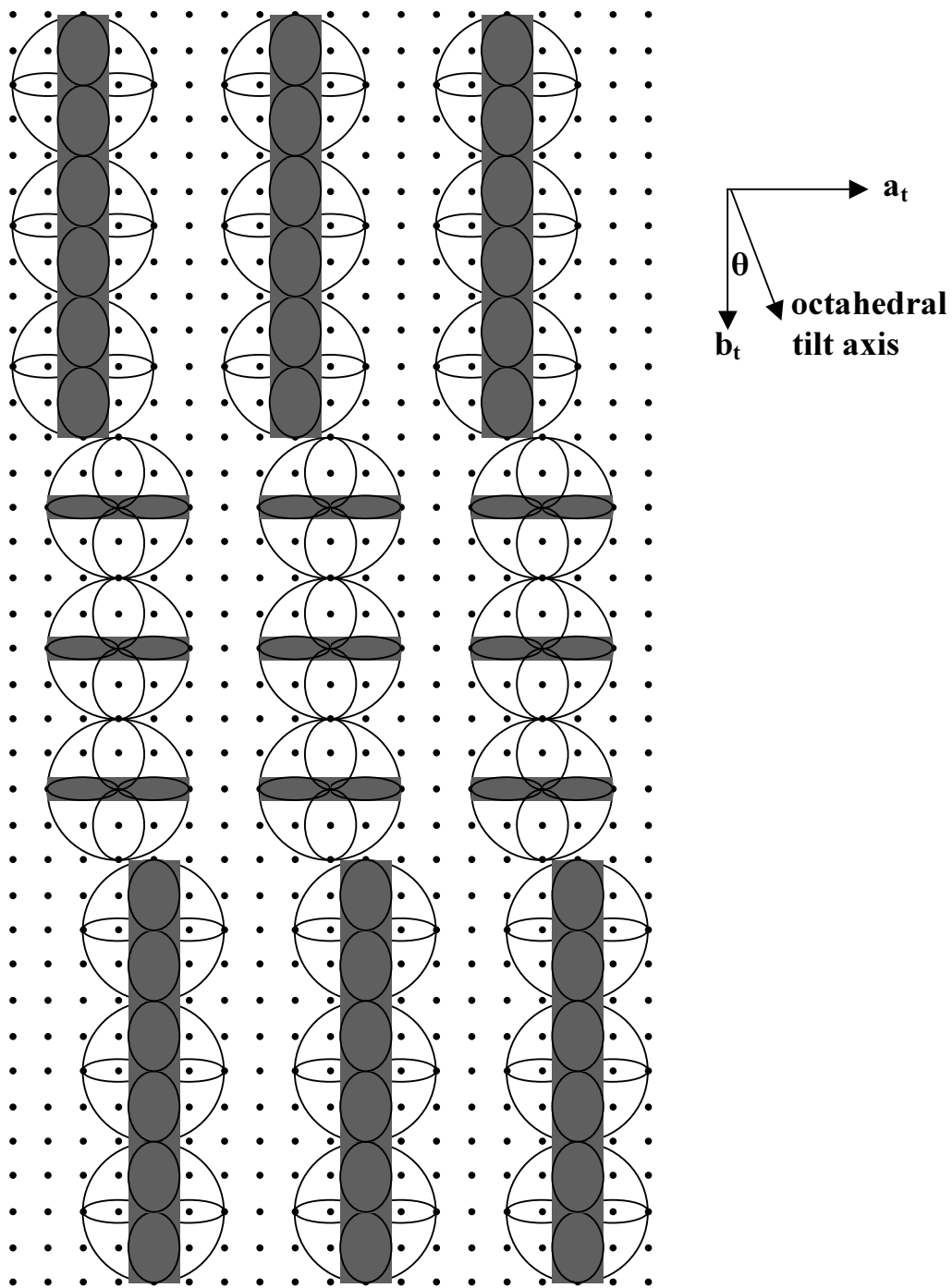

$a_t$

$\theta$

$b_t$

octahedral
tilt axis

Fig. 7   A. Rosencwaig

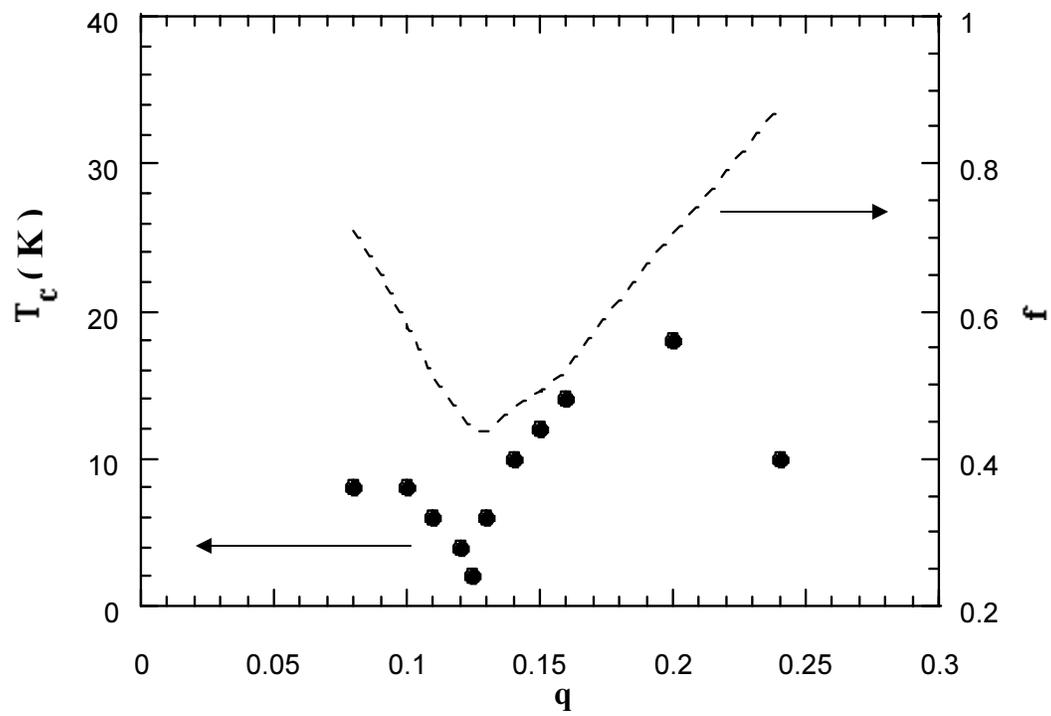

Fig. 8  A. Rosencwaig

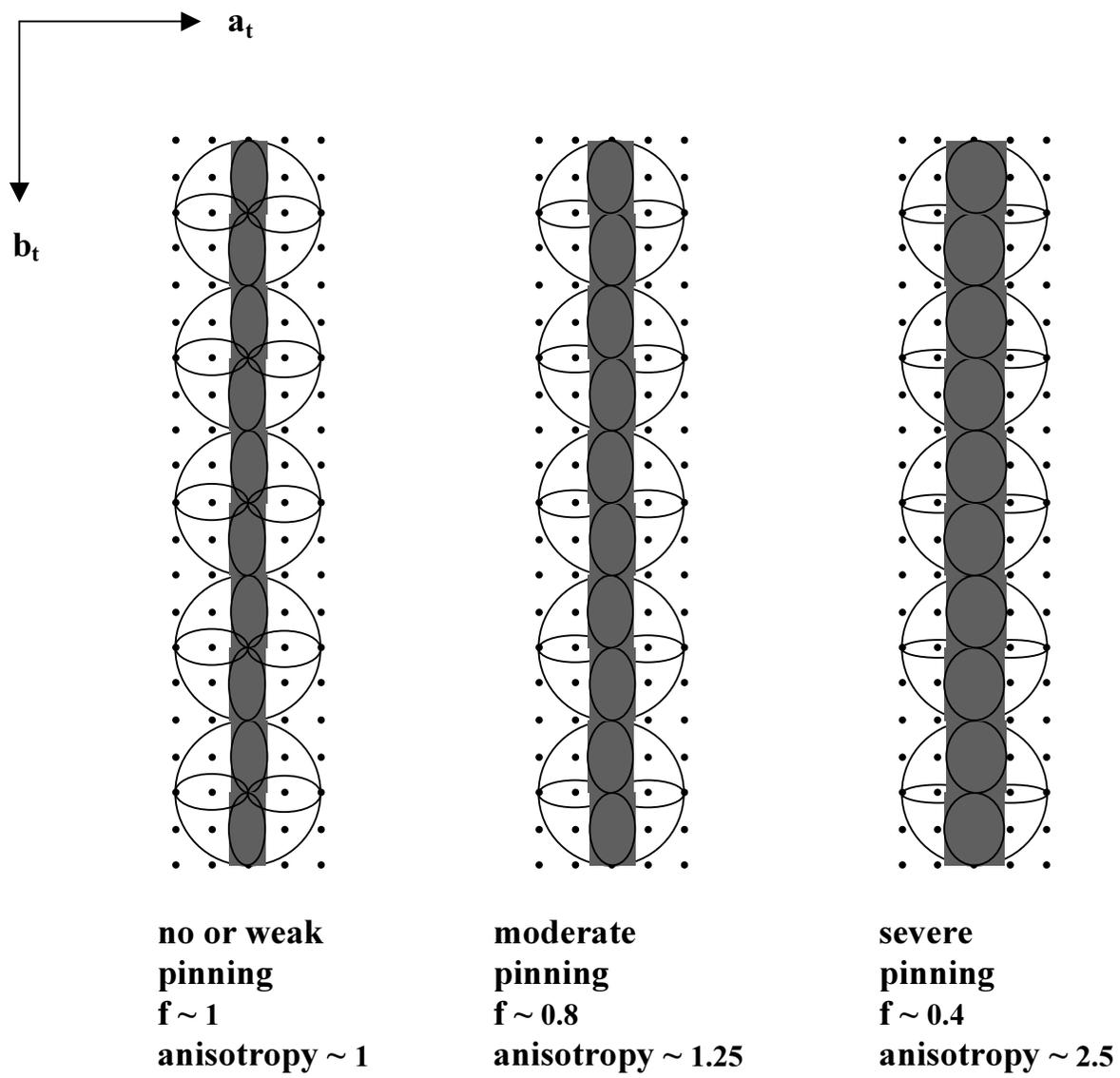

**a_t** $\mathbf{a_t}$

**b_t** $\mathbf{b_t}$

| no or weak pinning | moderate pinning | severe pinning |
|---|---|---|
| **f ~ 1** | **f ~ 0.8** | **f ~ 0.4** |
| **anisotropy ~ 1** | **anisotropy ~ 1.25** | **anisotropy ~ 2.5** |

Fig. 9  A. Rosencwaig

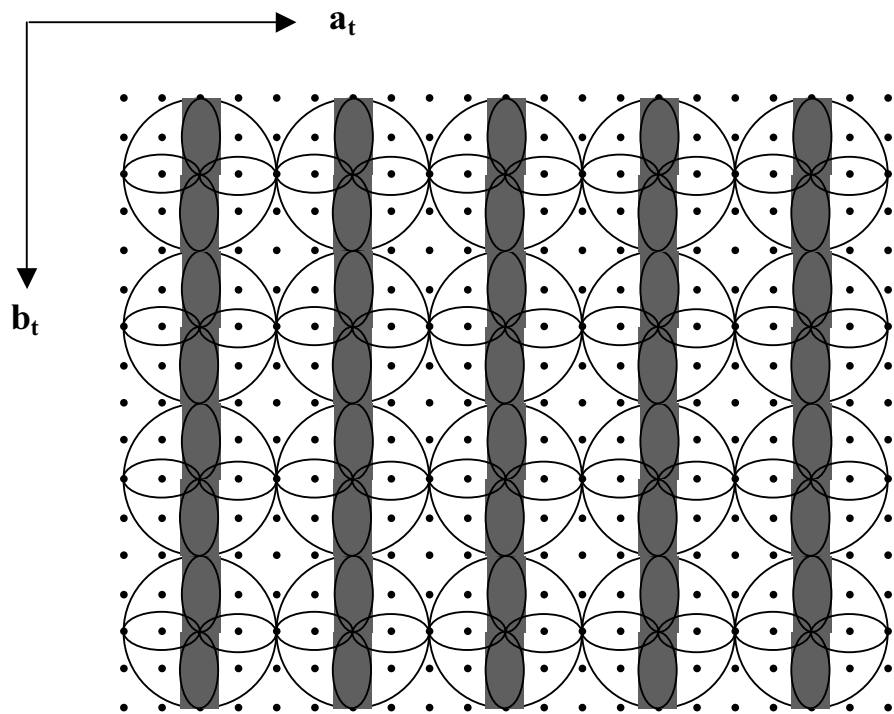

Fig. 10  A. Rosencwaig